\documentclass[pdflatex,sn-mathphys-ay]{sn-jnl}

\usepackage{graphicx}%
\usepackage{multirow}%
\usepackage{amsmath,amssymb,amsfonts}%
\usepackage{amsthm}%
\usepackage{mathrsfs}%
\usepackage[title]{appendix}%
\usepackage{xcolor}%
\usepackage{textcomp}%
\usepackage{manyfoot}%
\usepackage{booktabs}%
\usepackage{algorithm}%
\usepackage{algorithmicx}%
\usepackage{algpseudocode}%
\usepackage{listings}%
\usepackage{subcaption}

\theoremstyle{thmstyleone}%
\theoremstyle{thmstyletwo}%

\theoremstyle{thmstylethree}%
\usepackage{dsfont}
\newcommand{\cc}[1]{\mathcal{#1}}

\newcommand{\bb}[1]{\mathbb{#1}}
\newcommand{\bs}[1]{\boldsymbol{#1}}
\newcommand{\tht}{\theta}
\newcommand{\Tht}{\Theta}
\newcommand{\hth}{\hat{\theta}}
\newcommand{\lm}{\lambda}
\newcommand{\lms}{\lambda^*}

\newcommand{\Ns}{N^*}
\newcommand{\ns}{n^*}

\newcommand{\vep}{\varepsilon}

\newcommand{\vp}{\varphi}

\newcommand{\ind}{\mathds{1}}
\newcommand{\thh}{\textsuperscript{th }}

\newcommand{\til}[1]{\tilde{#1}}
\newcommand{\wh}[1]{\widehat{#1}}

\newcommand{\dd}{\mathrm{d}}

\DeclareMathOperator*{\siid}{\overset{iid}{\sim}}
\newcommand{\appropto}{\mathrel{\vcenter{
  \offinterlineskip\halign{\hfil$##$\cr
    \propto\cr\noalign{\kern2pt}\sim\cr\noalign{\kern-2pt}}}}}

\usepackage{csquotes}

\usepackage{algorithm}
\usepackage{hyperref}

\begin{document}

\title[Article Title]{\bf Parametric inference for the discretely observed multivariate Hawkes process using particle Markov Chain Monte Carlo}

\author*[1,2]{\fnm{Jason} \sur{Lambe}\orcid{https://orcid.org/0009-0000-8265-6043}}\email{\href{mailto:j.lambe@unsw.edu.au}{j.lambe@unsw.edu.au}}

\author[1]{\fnm{Feng} \sur{Chen}\orcid{https://orcid.org/0000-0002-9646-3338}}\email{\href{mailto:feng.chen@unsw.edu.au}{feng.chen@unsw.edu.au}}%

\author[1]{\fnm{Tom} \sur{Stindl}\orcid{https://orcid.org/0000-0001-8627-9337}}\email{\href{mailto:t.stindl@unsw.edu.au}{t.stindl@unsw.edu.au}}

\author[1]{\fnm{Tsz-Kit Jeffrey} \sur{Kwan}\orcid{https://orcid.org/0000-0003-2455-6348}}\email{\href{mailto:j.t.kwan@unsw.edu.au}{j.t.kwan@unsw.edu.au}}

\affil*[1]{\orgdiv{School of Mathematics and Statistics}, \orgname{UNSW Sydney}}


\affil[2]{\orgname{Defence Science and Technology Group}}


\abstract{The multivariate Hawkes process (MHP) is a useful statistical model for analysing multidimensional event time sequences that exhibit self-excitation and cross-excitation. When the MHP is monitored discretely, only the total number of events for each dimension in disjoint time intervals is observed. The likelihood function relative to this data is intractable, so traditional inference techniques are not available. To address this, we design an unbiased estimate of the intractable likelihood function using sequential Monte Carlo (SMC) based on a representation of the unobserved event times as latent variables in a state-space model. The unbiasedness of the SMC estimate allows for its use in place of the true likelihood in a Metropolis-Hastings algorithm, enabling the construction of a Markov Chain Monte Carlo sample from the posterior distribution over the parameters of the MHP. Using simulated data, we assess the performance of our method and demonstrate that it outperforms existing approaches in terms of mean squared error and computational efficiency. Terrorist activity in Afghanistan and Pakistan from 2018 to 2021 is analysed based on daily count data to examine the dynamics of terrorism in the region.
}



\maketitle

\section{Introduction}\label{sec:intro}

The Hawkes process \citep{hawkesSpectraSelfexcitingMutually1971} is a \textit{self-exciting} point process, meaning that the occurrence of an event causes a short-term spike in the arrival rate of subsequent events. The multivariate Hawkes process \citep[MHP;][]{hawkesSpectraSelfexcitingMutually1971} is a multidimensional extension of the Hawkes process in which events of finitely many \textit{types} can both self-excite events of the same type and \textit{cross-excite} events of other types. The MHP is a useful model in application domains where excitation is seen across multiple processes, such as terrorism and political violence \citep{tenchSpatiotemporalPatternsIED2016, junFlexibleMultivariateSpatiotemporal2024, zhouBayesianInferenceAggregated2025}, joint analysis of activity in a neuron cluster \citep{bonnetNeuronalNetworkInference2022}, financial market transactions during periods of positive or negative returns and market sentiment \citep{yangApplicationsMultivariateHawkes2018}, and the spread of infectious disease between cities or nations \citep{chenNovelPointProcess2022}.


The MHP model is often specified parametrically, with the goal of inference being to produce a point estimate with error quantification of the unknown parameter vector. The Maximum Likelihood Estimator (MLE) can be calculated for both the Hawkes process \citep{ogataAsymptoticBehaviourMaximum1978} and the MHP \citep{bowsherModellingSecurityMarket2007}. Expectation Maximisation (EM) algorithms are also commonly used to estimate the parameters of the MHP as they can mitigate the instability that may arise when numerically maximising the log-likelihood surface, which is often flat or multimodal \citep{veenEstimationSpaceTime2008}. However, both the MLE and the EM methods require knowledge of the precise event times over the span of observation. There are a variety of reasons that event time data may be aggregated for relevant applications of the MHP, such as the imprecise reporting of terror attacks \citep{tenchSpatiotemporalPatternsIED2016}, the inability of an affected individual to identify the time they became infected by a disease \citep{chenNovelPointProcess2022}, or where finite measurement precision makes it appear that financial transactions have occurred simultaneously \citep{bowsherModellingSecurityMarket2007}. With only discrete observation, the likelihood function of the MHP is analytically intractable, which poses a challenge for conducting inference.

Recent works have presented methods for fitting the MHP with aggregated data. An early method is given by \cite{kirchnerEstimationProcedureHawkes2017}, which approximates the likelihood of the discretely observed MHP by a sequence of order $p$ Integer-Valued Autoregressions (INAR), from which an estimator is obtained. This method is justified by a weak convergence of a sequence of INAR($\infty$) models to the MHP as the width of the observation intervals approaches zero. \cite{shlomovichMultivariate2022} use a modified Monte Carlo Expectation Maximisation (MCEM) algorithm to estimate the MHP with discretely observed data, with flexibility to handle differently sized observation periods. In the E-step of the algorithm, the authors sequentially reconstruct the latent event times by selecting the mode of a proposal distribution truncated over each observation period. This differs from typical MCEM methods as the hidden times are chosen deterministically instead of randomly sampled. The MCEM estimators appear to be biased in general, though simulation experiments in \cite{shlomovichMultivariate2022} demonstrate the superiority of this method over the INAR($p$) approach of \cite{kirchnerEstimationProcedureHawkes2017}. The estimation of standard errors of the MCEM estimates is not addressed in \cite{shlomovichMultivariate2022}.

\cite{koyamaCoarseGrainedHawkesProcesses2025} introduces the coarse-grained Hawkes process, which is an integer-valued time series that evolves according to a discretised version of the Hawkes intensity process. The discretisation scheme approximates relevant intractable integrals by assuming that the latent event times are uniformly distributed within each censoring interval, thereby capturing intra-bin excitation effects, albeit with some bias, which is minor for small levels of aggregation. When working with a stationary Hawkes process, the coarse-grained Hawkes process matches the original Hawkes process in terms of expected counts, and the second-order statistical properties are equal in the limit as the censoring interval width approaches $0$. The estimation procedure is a two-step process involving the minimisation of a loss function to obtain branching ratio and offspring kernel parameter estimates, followed by a method of moments estimation of the background rate. The coarse-grained Hawkes process therefore cannot handle time-varying background rates or unequal censoring intervals. Standard error estimation is also not addressed.

A Bayesian approach to the estimation problem is presented in \cite{zhouBayesianInferenceAggregated2025}, which also applies to settings with a spatial component. An expression for the likelihood is derived by treating the precise event times, locations, and branching structure as latent variables. Samples are drawn from the joint posterior over the parameters and latent variables via MCMC, with the marginal posterior over the parameters used to produce point estimates and credible intervals. Various algorithms for implementing the technique are provided, balancing computational efficiency against the autocorrelation of the posterior samples. The complete method is in general computationally heavy due to the need to evaluate the distribution of the parent index for each latent event, though this is mitigated in practice by artificially truncating the support of the offspring distribution.

In this work, we propose to obtain an unbiased estimator of the likelihood function via \textit{sequential Monte Carlo} (SMC) by extending the methods in \cite{chenEstimatingHawkesProcess2025}. The SMC estimate of the likelihood is then used in place of the true likelihood in an otherwise typical Metropolis-Hastings algorithm \citep{metropolisEquationStateCalculations1953, hastingsMonteCarloSampling1970} to construct a Markov chain for obtaining a dependent sample from the posterior distribution, a method commonly known as the Pseudo-Marginal Metropolis-Hastings \citep[PMMH;][]{andrieuPseudomarginalApproachEfficient2009} algorithm. The PMMH chain yields the true likelihood distribution (or more generally, the posterior distribution) as the stationary distribution of the resulting MCMC sample, from which the estimates are taken to be the mean or median. For this to be computationally feasible in the multivariate setting, we implement a highly efficient proposal distribution that guarantees that all particles agree with the observed count data. Numerical experiments will demonstrate the significant variance reduction in the SMC estimates compared to using a multivariate version of the Poisson proposals in \cite{chenEstimatingHawkesProcess2025}, which is crucial for ensuring the efficiency of the PMMH algorithm without requiring a prohibitively large number of particles. We implement the SMC algorithm using \textit{adaptive resampling} \citep{liuSequentialMonteCarlo1998} to achieve further variance reductions.


Since the PMMH estimates are unbiased and target the MLE, we see empirical consistency and strong performance in terms of mean-squared error relative to the available alternatives in the literature. Credible intervals are easily obtained by extracting the relevant quantiles of the MCMC sample. Importantly, the PMMH estimator is flexible to handle count data with unequally sized censoring intervals and Hawkes processes with non-stationary background rate functions, with little additional work. Due to the efficiency gains of our suggested proposal and the ease of parallelising SMC estimation, our method is competitive with high-quality alternatives in terms of speed.

The article is organised as follows. In Section \ref{sec:dat_mod_est}, we introduce the MHP model, formally set up the estimation problem, and present the estimation method. Simulation studies are performed in Section \ref{sec:sim_stdy} to compare the efficiency of the uniform proposal scheme against the Poisson proposals, and to assess the performance of the PMMH estimates across varying levels of aggregation. A comparison of the PMMH estimates to the MCEM \citep{shlomovichMultivariate2022}, coarse-grained Hawkes process \citep{koyamaCoarseGrainedHawkesProcesses2025}, and Bayesian MCMC \citep{zhouBayesianInferenceAggregated2025} estimation procedures will also be presented. In Section \ref{sec:appld_stdy}, the proposed method is used to estimate a bivariate Hawkes process model of terrorist activity in Afghanistan and Pakistan over the period of 2018--2021. The article concludes with a short discussion in Section \ref{sec:conc}. Detailed derivations and additional figures are provided in the appendices.

\section{Data, Model and Estimation Method}
\label{sec:dat_mod_est}
Let $\{(\tau_i, z_i)\}_{i\in\bb Z_+}$ denote a realisation of a multivariate point process. The sequence $\{\tau_i\}_{i\in\bb Z_+}$ is assumed to be strictly positive and strictly increasing, with $\tau_i$ interpreted as the $i$\thh event time after an initial time $t\, =\, 0$. For some $M\in \bb Z_+$, the value $z_i\in \{1,\, \dots,\, M\}\, =:\, \cc M$ represents the \textit{type} of the $i$\thh event. This structure gives rise to the $M$-dimensional counting process
\begin{align*}
	\bs N_t \ &= \ \big(N_1(t),\,\dots,\, N_M(t)\big),
\end{align*}
with $N_m(t)$ being the number of type-$m$ events that have occurred from the initial time $t=0$ up to and including time $t$. More generally, $N_m(A)$ is the number of type-$m$ events on a set $A\in \cc B(\bb R_+)$. The history of the process is represented by the filtration $\{\cc F_t\}_{t\geq 0}$, with $\cc F_t = \sigma\{\bs N_s: s \leq t\}$. To specify the model as an MHP, we use the conditional intensity vector $\bs \lm:\bb R_+  \to \bb R_+^M$, with $m$\thh component $\lm_m:\bb R_+  \to \bb R_+$ defined as
\begin{align*}
	\lm_m(t) \ := \ \frac{\bb E\big[\mathrm{d}N_m(t)\mid \cc F_{t-}\big]}{\mathrm{d}t} \ &= \ \nu_m(t) \ + \ \sum_{k: \tau_k<t}\eta_{m, z_k}h_{m, z_k}(t\, -\, \tau_k)\\ &=: \ \nu_m(t) \ + \ \vp_m(t), 
\end{align*}
where $\cc F_{t-} = \sigma\{\bs N_s: s < t\}$. The function $\nu_m: \bb R_+ \to \bb R_+$, $m\in\cc M$, determines the background arrival rate for type-$m$ events. The \emph{excitation kernel} functions $h_{m, j}: \bb R_+ \to \bb R_+$, $m,j\in\cc M$, are probability density functions, which control the shape of the excitation effect of type-$j$ events on type-$m$ events' intensity. They are also referred to as the \textit{offspring density functions} due to the cluster Poisson process interpretation of Hawkes processes \citep{hawkesClusterProcessRepresentation1974}.  The offspring density function $h_{m,j}$ determines the birth time distribution of generation-1 offspring events of type-$m$ due to a type-$j$ event, given at least one such event.  The \textit{branching ratios} $\eta_{m,j}$, $m, j\in \cc M$, are positive constants that 
represent the respective expected numbers of generation-1 offspring events of type-$m$ due to a type-$j$ event. Together, $\eta_{m,j}h_{m, j}(\cdot)$ is called the \emph{excitation function}, henceforth written as $\psi_{m, j}(\cdot)$. Let $\eta=(\eta_{m,j})_{m,j\in\cc M}$ be the matrix of branching ratios. To ensure the stability of the model, it is assumed that $\rho(\eta)  <  1$, with $\rho(X)$ denoting the spectral radius of the matrix $X$. With this stability condition, in the case where all background rate functions $\nu_m(\cdot)$ satisfy $\nu_m(t) \equiv\nu_m \in \bb R_+$, the mean arrival rates of events of different types approach finite values given by 
\begin{align*} 
   \lim_{t\to\infty} \bb E\big([\lambda_1(t),\dotsc,\, \lambda_M(t)]^\top\big) \ = \ (\cc I-\eta)^{-1}(\nu_1,\dotsc,\, \nu_M)^\top,
\end{align*}
with $\cc I$ denoting the $M\times M$ identity matrix. It is further assumed that $h_{m,j}$ depends on a vector of parameters, $\tht_{m,j}$, with the vector of all parameters (including values of $\nu_m$ and $\eta_{m,j}$) labelled as $\tht$. The explicit dependence of $\bs\lambda$ on $\tht$ is suppressed for notational convenience.

The MHP induces the \textit{total process} $\Ns$, defined by $\Ns(t)\, =\, \sum_{m=1}^M N_m(t)$, which counts events of all types from the initial time up to and including time $t$. The intensity process of $\Ns$, referred to as the \textit{total intensity}, is given by
\begin{align*}
    \lms(t) \ &= \ \sum_{m=1}^M\lm_m(t) \ = \  \nu(t) \ + \ \sum_{m=1}^M\vp_m(t)
\end{align*}
where $\nu(t)\, :=\, \sum_{m=1}^M \nu_m(t)$ is the total background arrival rate. For some indexed time $t_i \in \bb R_+$, we use the notation $\Ns_i = \Ns(t_i)$.

\subsection{Data and Likelihood}
The log-likelihood for an MHP observed on the interval $(0, T]$, when the precise event times are known, has the form
\begin{equation*}
\log L_{\mathrm{c}}(\theta) \ = \ \sum_{k:\tau_k\, <\, T}\log \lm_{z_k}(\tau_k) \ + \ \int_0^{T}\lms(t)\mathrm{d}t,
\end{equation*}
see for example \cite{daleyIntroductionTheoryPoint2003a}. However, when only aggregated data is observed for each event type at fixed time points, the likelihood function becomes intractable and we must resort to an approximation of the likelihood. 
Let there be $I\in \bb Z_+$ observation times, satisfying $0 = t_0 < t_1 < \dots < t_I  = T$. At each time $t_i$, the vector
\begin{align*}
	\bs n_i\  =\ (n_{i,1},\,\dots,\, n_{i,M})
\end{align*}
is observed, where $n_{i,m}$ represents the realised value of $N_m(t_{i-1}, t_i]$. The likelihood function is then given by
\begin{align*}
	L(\tht) \ &= \ P_\tht\big(N_m(t_{i-1},\, t_i]\, =\, n_{i,m}, \, m=1,\dotsc, M,\, i\, =\, 1,\,\dots,\, I\big).
\end{align*}
We henceforth write $P_\tht\big(N_m(t_{i-1}, t_i] = n_{i,m}\big)$ as $p_\tht(n_{i,m})$, and we make use of the colon notation for a collection of indexed variables or values, such that $x_{k:l} = (x_k, \dotsc,x_l)$ for $k < l$.

\subsection{Sequential Monte Carlo estimation of the likelihood}\label{sec:smc_mtd}
SMC enables sampling from the sequence of \textit{filtering} distributions $P^{(i)}_\tht(\mathrm{d}(\tau, z)_{1:\Ns_i}\mid \bs n_{1:i})$, with a sample at time $t_i$ obtained from the existing sample via a resampling and proposal scheme. An unbiased estimate of the likelihood is derived using the factorisation $L(\tht) = \prod_{i=1}^I p_\tht(\bs n_i\mid \bs n_{1:i-1})$. The discretely observed MHP admits a convenient state space representation. Let $x_i = (\tau, z)_{\Ns_{i-1}+1:\Ns_i}$ be the latent event times and types in the $i$\thh observation window, and let the total number of events on the $i$\thh window be $n^*_i = \sum_{m=1}^M n_{i, m}$. Since each $x_i$ contains both continuous and discrete variables, we utilise the product measure $\mu(\dd x_i) = \kappa_{\cc M}(\dd z_{\Ns_{i-1}+1:\Ns_i})\dd \tau_{\Ns_{i-1}+1:\Ns_i}$ as the reference measure when describing its distribution, where $\kappa_{\cc M}$ is the counting measure on $\cc M^{n^*_i}$. The transitions of the latent sequence are governed by the distribution of the MHP,
\begin{equation*}
    P_\tht(\dd x_i\mid x_{1:i-1}) \ = \ \prod_{k=\Ns_{i-1}+1}^{\Ns_i}\lm_{z_k}\big(\tau_k\big)\, \exp\Big(-\int^{\tau_{\Ns_i}\vee t_{i-1}}_{t_{i-1}}\lms(t) \dd t\Big) \mu(\dd x_i).
\end{equation*}
The observation mechanism defines the distribution of the signal $\bs n_i$, conditional on the latent event sequence. By standard results pertaining to the survival time for point processes \citep{daleyIntroductionTheoryPoint2003a}, we have that
\begin{equation*}
    p_\tht(\bs n_i \mid x_{1:i}) \ = \ \exp\Big(-\int^{t_i}_{\tau_{\Ns_i}\vee t_{i-1}}\lms(t) \dd t\Big).
\end{equation*}
Now, let $\{Q^{(i)}_\tht(\dd x_i \mid x_{1:i-1}, \bs n_{1:i})\}_{i\in \bb Z_+}$ be a sequence of proposal distributions. The proposal distribution at the $i$\thh interval conditions on $\bs n_{1:i}$, allowing our observation of the signal to influence the proposal of event times and types. The Feynman-Kac representation of the sequence of filtering distributions is
\begin{equation}\label{eq:FK}
    P^{(i)}_\tht(\dd x_{1:i}\mid \bs n_{1:i}) \ = \ \frac{1}{p_\tht(\bs n_{1:i})}\prod_{j=1}^i G_j(x_{1:j}, \bs n_{1:j}) Q^{(j)}_\tht(\dd x_j \mid x_{1:j-1}, \bs n_{1:j}),
\end{equation}
where $\{G_i\}_{i\in \bb Z_+}$ is the sequence of weight functions defined by
\begin{align}
    G_i(x_{1:i}, \bs n_{1:i}) \ &= \ \frac{P_\tht(\dd x_i\mid x_{1:i-1}) p_\tht(\bs n_i \mid x_{1:i})}{Q^{(i)}_\tht(\dd x_i \mid x_{1:i-1}, \bs n_{1:i})}\nonumber\\
    &= \ \frac{\prod_{k=\Ns_{i-1}+1}^{\Ns_i}\lm_{z_k}\big(\tau_k\big)\, \exp\Big(-\int^{t_i}_{t_{i-1}}\lms(t) \dd t\Big)\mu(\dd x_i)}{Q^{(i)}_\tht(\dd x_i \mid x_{1:i-1}, \bs n_{1:i})}.\label{eq:G}
\end{align}

The SMC algorithm is based on the forward recursion
\begin{align}\label{eq:forward_rec_mhp}
    P^{(i)}_\tht(\dd x_{1:i}\mid \bs n_{1:i}) \ &= \ \frac{p_\tht(\bs n_{1:i-1})}{p_\tht(\bs n_{1:i})}G_i(x_{1:i}, \bs n_{1:i}) Q^{(i)}_\tht(\dd x_i\mid x_{1:i-1}, \bs n_{1:i})\nonumber \\&\hspace{3cm} \times \ P^{(i-1)}_\tht(\dd x_{1:i-1}\mid \bs n_{1:i-1}), 
\end{align}
which is derived directly from \eqref{eq:FK}. A rigorous exposition in terms of Feynman-Kac operators is presented in \cite{delmoralFeynmanKacFormulaeGenealogical2004}. The SMC algorithm proceeds by applying the \eqref{eq:forward_rec_mhp} to Monte Carlo approximations of the integrals. 

Let there be $J$ \textit{particles}, with a particle referring to a proposed sample path for use in Monte Carlo estimation. In the first observation window, the Feynman-Kac representation is
\begin{equation*}
    P^{(1)}_\tht(\dd x_1\mid \bs n_1) \ = \ \frac{1}{p_\tht(\bs n_1)}G_1(x_1, \bs n_1) Q^{(1)}_\tht(\dd x_1\mid \bs n_1).
\end{equation*}
A sample from $P^{(1)}_\tht$ is drawn by sampling $x_1^{(1:J)}\siid Q^{(1)}_\tht(\dd x_1\mid \bs n_1)$ and computing the importance weights $G_1^{(j)} = G_1(x_1^{(j)}, \bs n_1)$. The normalised weights are denoted by $W^{(j)}_1 \propto G_1^{(j)}$, with the estimate
\begin{equation*}
    \hat p_\tht(\bs n_1) \ = \ \frac{1}{J}\sum_{j=1}^J G^{(j)}_1.
\end{equation*}
Suppose at observation window $i-1$ we have the weighted sample $\{W_{i-1}^{(j)}, x_{1:i-1}^{(j)}\}_{j=1}^J$. To obtain a sample from $P^{(i)}_\tht$, we can recycle the current sample by applying the forward recursion in \eqref{eq:forward_rec_mhp}. First, we draw a sample $\til x_{1:i-1}\siid \hat P^{(i-1)}_\tht$, which is achieved by sampling the index set $1:J$ from the multinomial distribution with probabilities $W_{i-1}^{(1:J)}$. This is termed the \textit{resampling} step. For each $j = 1,\dotsc,J$, the resampled particles have normalised weights $\til W_{i-1}^{(j)} = \frac{1}{J}$. 

The following step is \textit{propagation}. For each $j = 1,\dotsc,J$, we draw
\begin{equation*}
    x_i^{(j)} \ \sim \ Q^{(i)}_\tht(\dd x_i \mid \til x_{1:i-1}^{(j)}, \bs n_{1:i}).
\end{equation*}
The final particle is the vector $x_{1:i}^{(j)} = (\til x_{1:i-1}^{(j)}, x_i^{(j)})$, with normalised weight $W_i^{(j)}\propto G_k^{(j)}$. An estimate of the likelihood factor $p_\tht(\bs n_i\mid \bs n_{1:i-1})$ is obtained through the Monte Carlo approximation
\begin{align*}
    p_\tht(\bs n_i\mid \bs n_{1:i-1}) \ &= \ \int G_i(x_{1:i}, \bs n_{1:i})Q_\tht^{(i)}(\dd x_i\mid x_{1:i-1},\bs n_{1:i}) P^{(i-1)}_\tht(\dd x_{1:i-1}\mid \bs n_{1:i-1})\\
    &\approx \ \sum_{j=1}^J \til W_{i-1}^{(j)}G_i^{(j)}\\
    &= \ \hat p_\tht(\bs n_i\mid \bs n_{1:i-1}).
\end{align*}

The resampling step ensures that at each observation window, only high-weight particles are propagated to the next window, which avoids particle degeneracy. This typically performs far better than sequential importance sampling, which, after propagation through many observation windows, will often have only a single high-weight particle. However, resampling introduces an additional Monte Carlo noise into the likelihood estimates. The variance of our likelihood scheme can therefore be reduced by using \textit{adaptive resampling} \citep{chopinIntroductionSequentialMonte2020}, where at a given time window, resampling only takes place when the quality of the sample is determined to be dropping below an acceptable level. The statistical quality of the sample can be monitored via the \textit{effective sample size} \citep[ESS;][]{liuSequentialMonteCarlo1998}. The ESS is defined by
\begin{align*}
    \mathrm{ESS}\big(W^{(1:J)}_i\big) \ &= \ \frac{1}{\sum_{j=1}^J(W^{(j)}_i)^2}.
\end{align*}
An ESS of $J$ corresponds to the case of $W^{(j)}_i = \frac{1}{J}$ for all $j = 1,\dotsc, J$, which is characteristic of an independently drawn sample. Particle degeneracy occurs when $\mathrm{ESS} = 1$, as $W^{(j)}_i = 1$ for some $j = 1,\dotsc, J$, with the rest equal to $0$. For adaptive resampling, one chooses to resample only if $\mathrm{ESS}\big(W_i^{(1:J)}\big) < \mathrm{ESS}^*$, for some cutoff $ESS^*\in [1, J]$ \citep{delmoralSequentialMonteCarlo2006}. When no resampling occurs, we simply set $\til x^{(j)}_{1:i-1} = x^{(j)}_{1:i-1}$ and $\til W_{i-1}^{(1:J)} = W_{i-1}^{(1:J)}$. The complete SMC procedure with adaptive resampling is presented in Algorithm~\ref{alg:smc}. 

\begin{algorithm}
\caption{Sequential Monte Carlo with Adaptive Resampling}
\label{alg:smc}
\begin{algorithmic}[1] 
\Require Data $\bs n_{1:I}$, parameter $\tht\in \Tht$
\Ensure Likelihood estimate $\hat L(\tht)$

\State Initialise weights $\til W_0^{(j)} \gets 1/J$, particles $\til x_0^{(j)} = \{\}$ and likelihood estimate $\hat L(\tht) = 1$

\For{$i = 1$ to $I$}
    \For{$j = 1$ to $J$}
        \State Sample $x_i^{(j)} \sim Q^{(i)}_\tht(\dd x_i \mid \til x_{1:i-1}^{(j)}, \bs n_{1:i})$
        \State Compute weight increment $G_i^{(j)}$ 
        \State Update normalised weight $W_i^{(j)}\propto \til W^{(j)}_{i-1}G_i^{(j)}$
    \EndFor
    \State Compute likelihood factor estimate $\hat p_\theta(\bs n_i \mid \bs n_{1:i-1}) = \sum_{j=1}^J \tilde W_{i-1}^{(j)} G_i^{(j)}$
        \State Update likelihood estimate $\hat L(\tht) \gets \hat L(\tht) \times \hat p_\theta(\bs n_i \mid \bs n_{1:i-1})$
    \If{$\mathrm{ESS}\big(W^{(1:J)}_i\big) < \mathrm{ESS}^*$}
        \State Resample $\til x_{i}^{(1:J)}\, \siid\, \hat P^{(i-1)}_\tht(\dd x_{1:i}\mid \bs n_{1:i})$
        \State Set $\til W_i^{(j)} \gets 1/J$ for all $j = 1,\dotsc, J$
        \Else
        \State Set $\til x_{i}^{(1:J)} \gets x_{i}^{(1:J)}$ and $\til W_i^{(1:J)}\gets W_i^{(1:J)}$
    \EndIf
\EndFor

\end{algorithmic}
\end{algorithm}

Details on the computation of the numerator of $G_i(x_{1:i}, \bs n_{1:i})$ are provided in Appendix~\ref{app:A1_lw}. Importantly, in the case of an exponential offspring kernel, the cumulative excitation component of the intensity is Markovian \citep{oakesMarkovianSelfexcitingProcess1975a}, allowing for the computation of $G_i$ to be linear in the number of events and eliminating the need to store the complete history of each particle chain. Details are found in Appendix~\ref{app:A2_exp}.

The variance of the SMC likelihood estimate is heavily impacted by the choice of proposal distributions, $\{Q^{(i)}_\tht\}_{i\in \bb Z_+}$. The proposal $Q^{(i)}$ is required only to dominate the filtering distribution, $P^{(i)}_\tht$. In this work, we propose the latent event times on a given window from the ordered uniform distribution. Formally, let $\cc T_i = \{t_{i-1} < \tau_{\Ns_{i-1}+1} < \dotsc < \tau_{\Ns_i} \leq t_i\}$, which defines the set of valid event times on $(t_{i-1}, t_i]$. The proposal distribution for event times is given by
\begin{equation*}
    Q^{(i)}_\tht (\dd \tau_{\Ns_{i-1} + 1:\Ns_i}\mid \bs n_i) \ = \ \frac{n_i^*!}{(t_i - t_{i-1})^{n_i^*}}\, \ind_{\cc T_i}(\tau_{N_{i-1}^* + 1:\Ns_i}) \dd \tau_{N_{i-1}^* + 1:\Ns_i}.
\end{equation*}
This distribution is appealing both due to the simplicity of its expression, being the same for each particle, and for its statistical efficiency. The latter will be explored in Section~\ref{sec:unif_vs_poi}. For the latent event types, we have $n_i^*$ total events, with the knowledge that $n_{i, m}$ must be of type $m$, for all $m = 1,\dotsc, M$. We simply randomly assign each proposed event a type subject to the event counts of different types, which is equivalent to a multivariate hypergeometric distribution, that is
\begin{equation*}
    Q^{(i)}_\tht(\dd z_{\Ns_{i-1} + 1:\Ns_i}\mid \bs n_{i}) \ = \ \frac{\prod_{m=1}^M n_{i,m}!}{n^*_i!}\kappa_{\cc M}(\dd z_{\Ns_{i-1} + 1:\Ns_i}).
\end{equation*}
By proposing event times independently from event types, our proposal distribution is
\begin{equation*}
    Q_\tht^{(i)}(\dd x_{i}\mid \bs n_i) \ = \ \frac{\prod_{m=1}^M n_{i,m}!}{(t_i - t_{i-1})^{n_i^*}}\, \ind_{\cc T_i}(\tau_{N_{i-1}^* + 1:\Ns_i}) \mu(\dd x_i).
\end{equation*}
Though $Q^{(i)}_\tht$ is allowed to depend on the full history $x_{1:i-1}$ and the full observation sequence $\bs n_{1:i}$, we find our proposal to be simple and efficient in implementation. The uniform placement of event times conditional on the counts $\bs n_i$ is akin to approximating the evolution of the latent Hawkes process over $(t_{i-1}, t_i]$ by a Poisson process, which works exceptionally well (Section~\ref{sec:unif_vs_poi}).

\subsection{Parameter Estimation}
We will now provide some rationale for the estimation procedure. For more details, see \cite{chenEstimatingHawkesProcess2025}. Assuming that the log-likelihood function is sufficiently regular, then for a large number of intervals, $I$, we have the approximation
\begin{align*}
    \log L(\tht) \ &= \ \log p_{\tht}(\bs n_{1:I}) \ \approx \ \log p_{\hth}(\bs n_{1:I}) \ - \ \frac{1}{2}(\tht\, -\, \hth)^\top\big(-H(\hth)\big)(\tht\, -\, \hth),
\end{align*}
where $H(\tht)$ is the Hessian of the log-likelihood and $\hth$ is the MLE. Therefore, the likelihood function satisfies an approximate proportionality relation of the form
\begin{align*}
    L(\tht) \ \appropto \  \exp\Big(\frac{1}{2}(\tht\, -\, \hth)^\top\big(-H(\hth)\big)(\tht\, -\, \hth)\Big),
\end{align*}
which is a Gaussian density up to a normalising constant. It follows that, if the prior distribution of $\theta$ is not very informative and is fixed when the number of observation intervals $I$ tends to infinity, the posterior distribution will be asymptotically Gaussian, with mean $\hth$ and covariance matrix $H^{-1}(\hth)$. If the likelihood were explicitly available, an approximation to the MLE and Hessian could be obtained by using Markov Chain Monte Carlo (MCMC) to sample from $L(\tht)$, then taking the mean and variance of the resulting sample.

The Metropolis-Hastings algorithm is an appropriate method for constructing a Markov chain to approximate $L(\tht)$, as the likelihood is only known up to a constant of proportionality. Given the current state of the chain is $\tht$, we pick the proposal distribution $Q(d\tht'\mid \tht)$ to be Gaussian with mean $\tht$ and variance $\delta$. This is symmetric and dominates the target distribution. Taking a draw $\tht'\sim Q(d\tht'\mid \tht)$, we accept $\tht'$ as the new state of the chain with probability $L(\tht')/L(\tht)$, else we reject and remain at $\tht$.

Since we cannot evaluate the likelihood function exactly when working with aggregated data, we instead use a PMMH algorithm \citep{andrieuPseudomarginalApproachEfficient2009}. This method functions identically to the typical Metropolis-Hastings method described, but replaces the likelihood $L(\tht)$ with the SMC estimate $\widehat{L}(\tht)$. This is justified due to the unbiasedness of the SMC estimator. It is important that, when implementing the algorithm, the value $\widehat{L}(\tht)$ is not recalculated each time the acceptance probability $L(\tht')/L(\tht)$ is calculated -- it should be done only once, with the value reused \citep{andrieuPseudomarginalApproachEfficient2009}. See \cite{pittPropertiesMarkovChain2012} or \cite{chenEstimatingHawkesProcess2025} for a proof that the stationary distribution of a Markov chain constructed using the PMMH method will yield the true posterior distribution as its stationary distribution. 


\section{Simulation Studies}\label{sec:sim_stdy}
In this section, numerical experiments are conducted to verify the accuracy of the proposed SMC procedure to approximate the likelihood. Then, we numerically compare the ordered uniform proposal distribution used in our method to the Poisson proposals used by \cite{chenEstimatingHawkesProcess2025}. The performance of the estimators derived from the PMMH algorithm is assessed on simulated sample paths, and compared with the MCEM estimator \citep{shlomovichMultivariate2022}, the coarse-grained Hawkes process method \citep{koyamaCoarseGrainedHawkesProcesses2025}, and the Bayesian MCMC estimator \citep{zhouBayesianInferenceAggregated2025}.

\subsection{SMC Estimate of the Likelihood} \label{ssec:SMC_ver}
\begin{figure}[ht]
\begin{subfigure}{0.49\textwidth}
\centering
    \includegraphics[width = \textwidth]{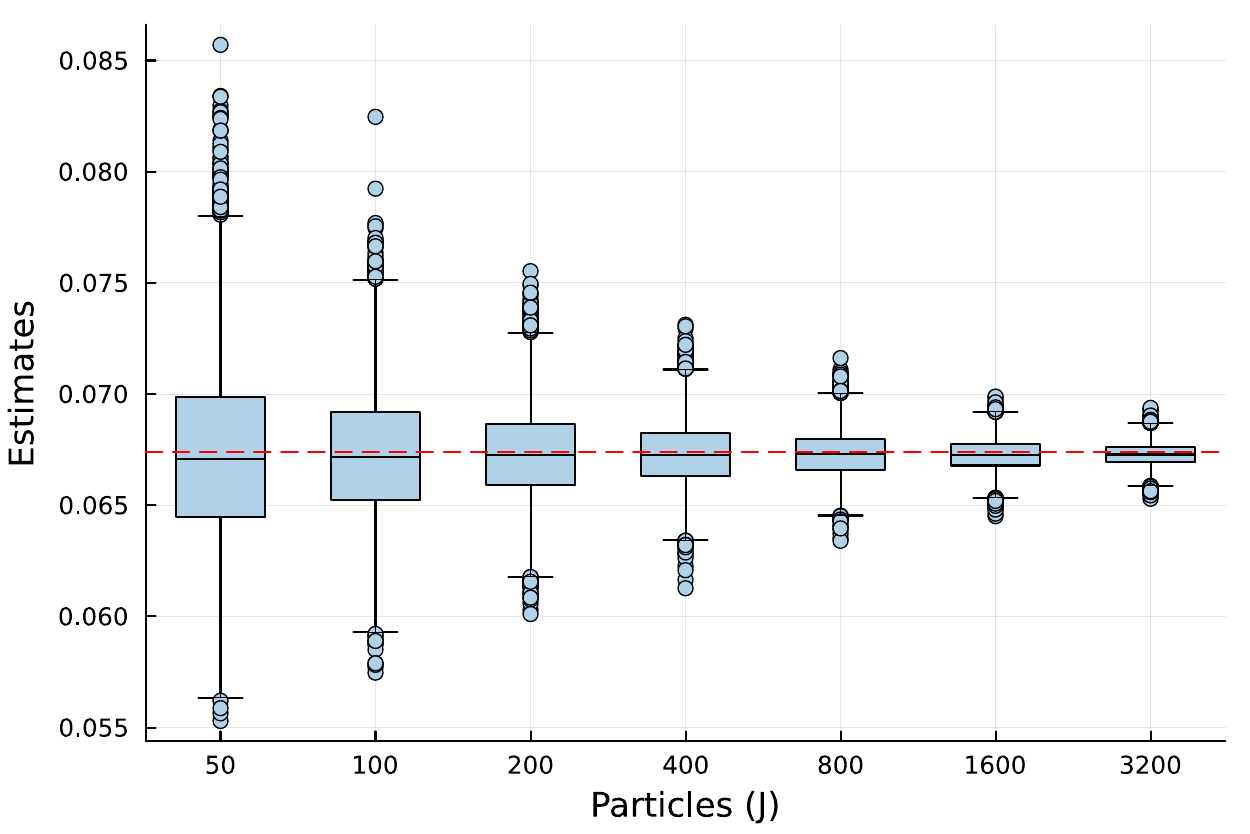}
    \caption{$P(A)$, MHP with exponential kernel.}
    \label{fig:MC_exp}
\end{subfigure}
\hfill
\begin{subfigure}{0.49\textwidth}
    \centering
    \includegraphics[width = \textwidth]{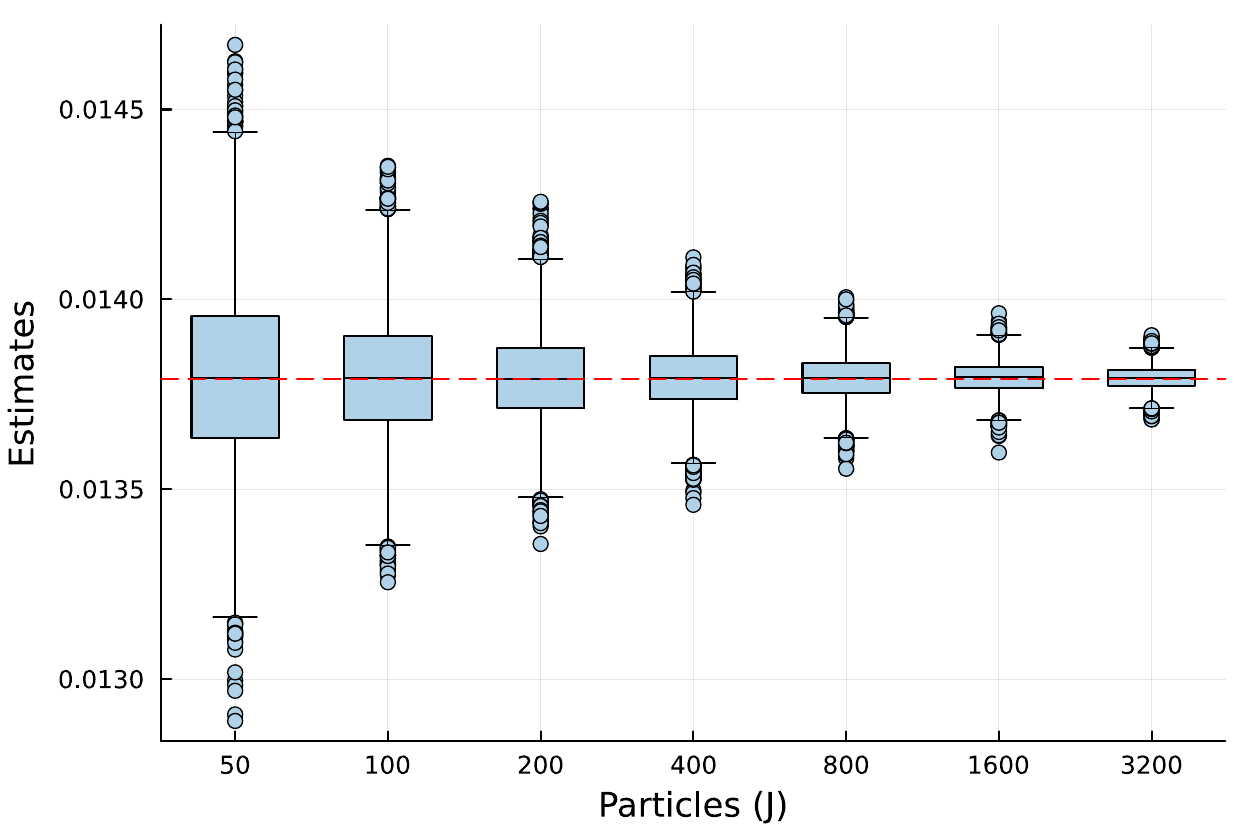}
    \caption{$P(B)$, MHP with gamma kernel.}
    \label{fig:MC_gam}
\end{subfigure}
    \caption{Box plots of $10,\!000$ SMC estimates of $P(A)$ and $P(B)$ respectively, with $A$ and $B$ defined in Section \ref{ssec:SMC_ver}. The Monte-Carlo estimates are marked by the red dashed lines.}
    \label{fig:boxplot_exp}
\end{figure}

We now provide two experiments to verify the validity of the SMC algorithm. First, consider a bivariate Hawkes process with exponential excitation kernels $h_{i,j}(t) = \tfrac{1}{\beta_{ij}}e^{-t/\beta_{ij}}$. The background intensity vector is $\nu = (1,\, 1)$, and the excitation parameters are $\beta_{ij} = 0.5$ for any combination of $i, j\in \{1, 2\}$. The branching ratio matrix $\eta$ has diagonal entries of $0.6$ and off-diagonal entries of $0.4$. Define the event $A := \big\{N_1(0, 1] = 1, N_2(0, 1] = 1\big\}$, that is, precisely one event of each type in the first unit of time. A standard Monte Carlo approach involving the simulation and inspection of $10^{6}$ samples yields the estimate $0.0674$ for $P(A)$. Figure~\ref{fig:MC_exp} displays $10,\!000$ SMC estimates of $P(A)$, with varying number of particle $J$. The mean estimate is close to the Monte Carlo estimate even for low $J$ values, with the empirical bias and variance decreasing in $J$.

Next, consider again a bivariate Hawkes process with baseline intensity vector $\nu$ and branching ratio matrix $\eta$ as before, but now characterised by gamma offspring densities $h_{i,j}(t) = t^{\kappa_{ij}-1} e^{-t/\delta_{ij}} / (\Gamma(\kappa_{ij}) \delta_{ij}^{\kappa_{ij}})$, with shape and scale parameters given respectively by
\begin{align*}
	\kappa \ &= \ \begin{pmatrix}
2 & 3\\
3 & 2
\end{pmatrix},\qquad \delta \ = \ \begin{pmatrix}
1 & 2\\
2 & 1
\end{pmatrix}.
\end{align*}
Define the event $B := \big\{N_1(0, 1] = N_1(1, 2]  =   N_2(0, 1] =N_2(1, 2]  = 1\big\}$. The standard Monte Carlo approach estimates $P(B)$ to be $0.0138$ using $10^6$ sample paths. Figure~\ref{fig:MC_gam} shows convergence of the mean SMC estimate of $P(B)$ to the Monte Carlo estimate, and reduction in the variance as the number of particles increases.

\subsection{Evaluation of the Proposal Distribution}\label{sec:unif_vs_poi}
Our proposed SMC algorithm implements an ordered uniform proposal distribution for sampling latent event times, which we now compare with the finite distributions of a Poisson process used in \cite{chenEstimatingHawkesProcess2025}. Consider a representative observation interval $(t_{i-1}, t_i]$, with unobserved event times $\tau_{\Ns_{i-1}+1:\Ns_i}$ and types $z_{\Ns_{i-1}+1:\Ns_i}$. The Poisson proposal method is designed such that proposed times $\tau_{\Ns_{i-1}+1:\Ns_i}$ are distributed according to the first $\ns_i$ events of a homogeneous Poisson process on $(t_{i-1}, \infty)$ with rate parameter $\rho = \gamma_{0.95; \ns_i, 1} / (t_i - t_{i-1})$, taking $\gamma_{0.95; \ns_i, 1}$ to be the $0.95$ quantile of the gamma distribution with shape parameter $\ns_i$ and rate $1$. \cite{chenEstimatingHawkesProcess2025} select $\gamma_{0.95; \ns_i, 1}$ as this achieves a probability of $0.95$ that the proposed event times will fall within the $i$\thh observation window. Event types are then assigned from the multinomial distribution, as described in Section \ref{sec:smc_mtd}. This is a natural multivariate extension of the univariate method of \cite{chenEstimatingHawkesProcess2025}. For the ordered uniform proposals, we begin by sampling the first $\ns_i + 1$ times of the Poisson process with rate $1$, followed by a simple transformation to return $\ns_i$ ordered uniforms on $(t_{i-1}, t_i]$. See sections~\textit{V.2} and~\textit{V.3} of \cite{devroyeNonuniformRandomVariate1986} for relevant proofs and discussion of computational complexity. The computational time of both methods is linear in the number of particles.

\begin{figure}[ht]
\centering
  \includegraphics[width = 0.75\linewidth]{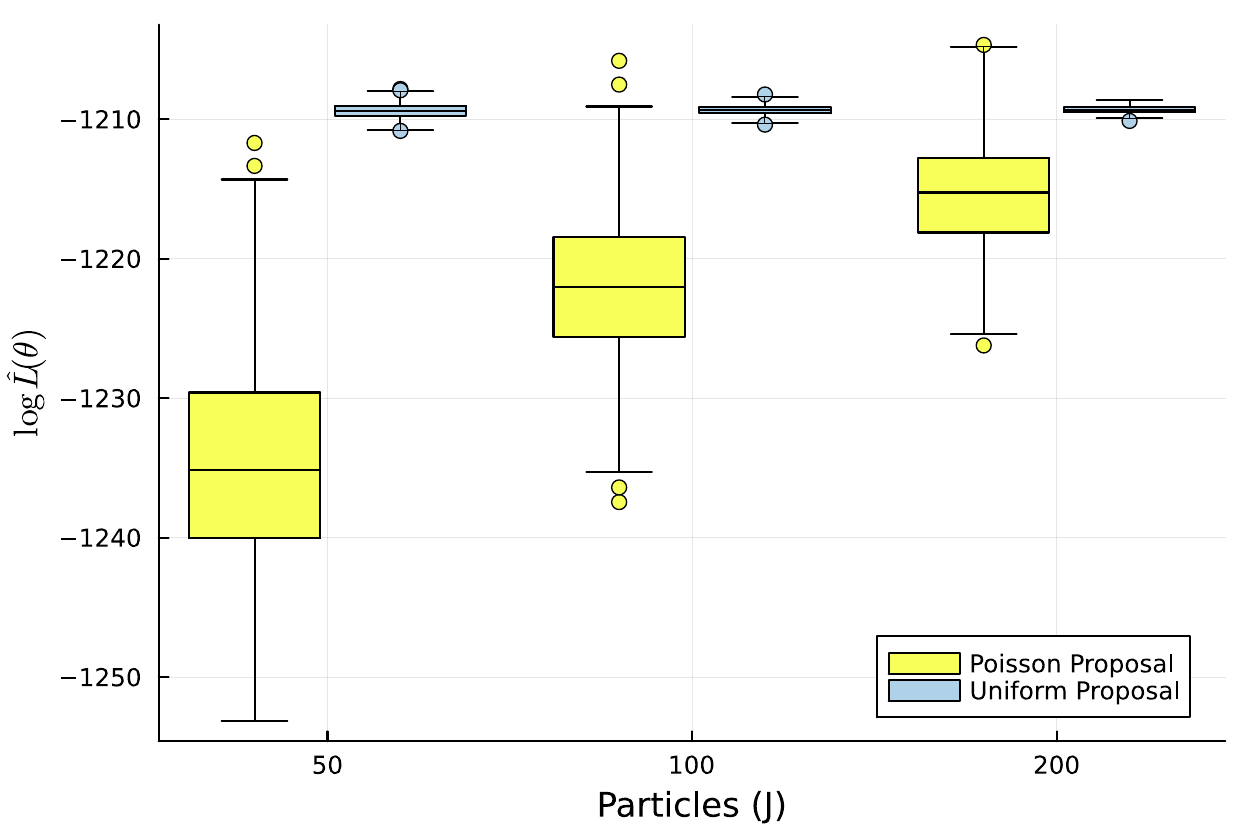}
  \caption{Box plots of $500$ SMC estimates of $\log L(\tht)$, using uniform and Poisson proposals.}
  \label{fig:boxplot_var_comp}
\end{figure}

One advantage of the ordered uniform distribution is that proposals are guaranteed to agree with observations, whereas with Poisson proposals, on average 5\% of the particles disagree with the observations, resulting in the assignment of zero weights and a loss of efficiency. The ordered uniform proposal method also yields significantly lower variance in the SMC likelihood estimates compared to using Poisson proposals. The magnitude of this improvement is illustrated in Figure~\ref{fig:boxplot_var_comp}, which shows 500 log-likelihood estimates at the true parameter on a single sample path with censoring time $T = 200$ and aggregation size $\Delta = 0.5$, for different numbers of particles. Such efficiency gains allow the PMMH chain to be run with fewer particles (by the order of $10^3$ or $10^4$) while maintaining accurate SMC likelihood estimates. Accurate SMC estimates reduce the chance that an unusually high likelihood estimate causes the PMMH chain to remain in place for extended periods. Given the increased demands of estimating many parameters in the multivariate setting, this efficiency gain is crucial for the practicability of the PMMH estimation method. 

\subsection{MLE Approximation via PMMH}\label{ssec:PMMH}
This section assesses the finite-sample performance of the PMMH-MCMC estimator of the model parameters. We simulate $S = 500$ sample paths up to time $T=800$ of the bivariate Hawkes process with exponential kernels and parameters
\begin{align*}
    \nu \ =\ \begin{pmatrix}
        0.8\\ 1.0
    \end{pmatrix}, 
    \quad \eta \ =\ \begin{pmatrix}
        0.6 & 0.3 \\ 0.25 & 0.5
    \end{pmatrix}, 
    \quad \beta \ = \ \begin{pmatrix}
        0.5 & 0.5 \\ 0.75 & 0.75
    \end{pmatrix}.
\end{align*}
The events are then aggregated into observation periods of fixed width $\Delta\in \{0.1, 0.5, 1.0\}$. The true parameter yields an average of approximately $10.5$ events per unit time. The PMMH algorithm is run for $10,\!000$ iterations, with the approximate MLE for each parameter taken to be the median of the resulting sample. The upper and lower $2.5$ percentile points define the approximate 95\% Wald confidence interval, with the width of the interval divided by $2\Phi^{-1}(0.975)$ to be the estimated standard error of the approximate MLE. Letting $n_1$ and $n_2$ be the total number of type 1 and type 2 events, respectively, each chain is initialised with $\nu_0 = \big(n_1/2T, n_2/2T\big)$ and the initial branching ratio $\eta_0$ has diagonal entries of $0.3$ and off-diagonals of $0.2$, which naively assumes that half of the observations are background events. All $\beta_{i,j}$ terms are initialised at $1$. 

The case of $\Delta = 0$ is also considered, corresponding to knowledge of exact event times, for which the log-likelihood is explicitly known. The MLE is therefore obtained by minimising the negative log-likelihood function, with estimates of the standard error obtained by taking the square roots of the diagonal entries of the inverse Hessian matrix. Estimates are also computed using only the data to a shortened censoring time of $T = 400$ to assess the improvement in estimation quality with additional data. The results are summarised in Table~\ref{tab:pmmh_sim_exp}, which reports the mean approximate MLE over the 500 simulated paths (Est), the empirical standard error of the 500 estimates (SE), the mean of the standard error estimates computed using the approximate confidence interval as previously described ($\widehat{\text{SE}}$) and the empirical coverage probability of the approximate confidence intervals (CP).

\begin{table}[t]
    \centering
        \caption{Summary of PMMH estimates for simulated sample paths}
    \label{tab:pmmh_sim_exp}
    \begin{tabular}{lcccccccccccc}\toprule
    &&&$\nu_1$ & $\nu_2$ & $\eta_{1,1}$  & $\eta_{1,2}$ & $\eta_{2,1}$ & $\eta_{2,2}$ & $\beta_{1,1}$ & $\beta_{1,2}$ &  $\beta_{2,1}$& $\beta_{2,2}$\\
    \midrule
     $\Delta$& $T$ && $0.8$ &$ 1.0$ & $0.6$ & $0.3$ & $0.25$ & $0.5$ & $0.5$ & $0.5$ & $0.75$ & $0.75$\\
    \midrule 
  \multirow{9}{*}{0} & \multirow{4}{*}{$400$} &  Est & 0.818 & 1.028 & 0.593 & 0.303 & 0.256 & 0.488 & 0.500 & 0.515 & 0.782 & 0.746\\
&&SE & 0.168 & 0.183 & 0.045 & 0.054 & 0.048 & 0.057 & 0.066 & 0.159 & 0.253 & 0.148\\
&&$\wh {\mathrm{SE}}$ & 0.165 & 0.178 & 0.042 & 0.051 & 0.045 & 0.057 & 0.063 & 0.147 & 0.235 & 0.145\\
&&CP & 0.946 & 0.942 & 0.934 & 0.938 & 0.938 & 0.952 & 0.936 & 0.918 & 0.928 & 0.930\\
\cmidrule(lr){2-13}
   & \multirow{4}{*}{$800$} & Est & 0.821 & 1.022 & 0.594 & 0.302 & 0.255 & 0.489 & 0.499 & 0.507 & 0.778 & 0.735\\
&& SE & 0.116 & 0.130 & 0.029 & 0.036 & 0.035 & 0.043 & 0.047 & 0.105 & 0.169 & 0.102\\
&& $\wh {\mathrm{SE}}$ & 0.117 & 0.126 & 0.030 & 0.037 & 0.032 & 0.041 & 0.045 & 0.103 & 0.168 & 0.102\\
&& CP & 0.944 & 0.946 & 0.950 & 0.940 & 0.926 & 0.936 & 0.924 & 0.954 & 0.934 & 0.932\\
   \midrule
    
    \multirow{9}{*}{$0.1$} & \multirow{4}{*}{$400$} & Est & 0.803 & 1.009 & 0.593 & 0.305 & 0.254 & 0.492 & 0.506 & 0.542 & 0.786 & 0.765\\
&& SE & 0.175 & 0.189 & 0.043 & 0.058 & 0.048 & 0.058 & 0.068 & 0.195 & 0.267 & 0.147\\
&& $\wh{\mathrm{SE}}$ & 0.165 & 0.180 & 0.042 & 0.052 & 0.045 & 0.057 & 0.065 & 0.170 & 0.256 & 0.149\\
&& CP & 0.934 & 0.940 & 0.934 & 0.899 & 0.914 & 0.932 & 0.945 & 0.914 & 0.930 & 0.940 \\
   \cmidrule(lr){2-13}
   & \multirow{4}{*}{$800$} & Est & 0.800 & 1.014 & 0.595 & 0.306 & 0.252 & 0.494 & 0.500 & 0.521 & 0.762 & 0.752\\
&& SE & 0.123 & 0.128 & 0.029 & 0.039 & 0.034 & 0.042 & 0.045 & 0.116 & 0.185 & 0.108\\
&& $\wh{\mathrm{SE}}$ & 0.118 & 0.126 & 0.030 & 0.037 & 0.032 & 0.040 & 0.045 & 0.113 & 0.173 & 0.103\\
&& CP & 0.941 & 0.932 & 0.947 & 0.936 & 0.930 & 0.926 & 0.959 & 0.947 & 0.918 & 0.949\\
\midrule
\multirow{9}{*}{$0.5$} & \multirow{4}{*}{$400$}& Est & 0.806 & 1.005 & 0.597 & 0.299 & 0.253 & 0.495 & 0.512 & 0.545 & 0.767 & 0.768\\
&& SE & 0.179 & 0.187 & 0.046 & 0.063 & 0.050 & 0.059 & 0.076 & 0.256 & 0.308 & 0.158\\
&& $\wh{\mathrm{SE}}$ & 0.171 & 0.181 & 0.043 & 0.053 & 0.044 & 0.056 & 0.071 & 0.248 & 0.314 & 0.159\\
&& CP & 0.942 & 0.928 & 0.934 & 0.900 & 0.886 & 0.922 & 0.944 & 0.948 & 0.942 & 0.928\\

\cmidrule(lr){2-13}
& \multirow{4}{*}{$800$}& Est & 0.798 & 1.013 & 0.595 & 0.305 & 0.252 & 0.494 & 0.501 & 0.527 & 0.761 & 0.753\\
&& SE & 0.124 & 0.131 & 0.030 & 0.040 & 0.034 & 0.043 & 0.050 & 0.147 & 0.213 & 0.116\\
&& $\wh{\mathrm{SE}}$ & 0.120 & 0.127 & 0.030 & 0.037 & 0.032 & 0.040 & 0.050 & 0.151 & 0.204 & 0.110\\
&& CP & 0.946 & 0.940 & 0.946 & 0.916 & 0.934 & 0.932 & 0.948 & 0.946 & 0.934 & 0.956\\

\midrule
\multirow{9}{*}{$1.0$} & \multirow{4}{*}{$400$} & Est & 0.801 & 0.991 & 0.600 & 0.297 & 0.258 & 0.492 & 0.512 & 0.619 & 0.758 & 0.773\\
&& SE & 0.197 & 0.191 & 0.051 & 0.073 & 0.054 & 0.064 & 0.089 & 0.415 & 0.398 & 0.185\\
&& $\wh{\mathrm{SE}}$ & 0.172 & 0.180 & 0.045 & 0.058 & 0.046 & 0.058 & 0.080 & 0.291 & 0.303 & 0.169\\
&& CP &0.910 & 0.932 & 0.922 & 0.890 & 0.890 & 0.916 & 0.922 & 0.868 & 0.832 & 0.922\\
\cmidrule(lr){2-13}
   & \multirow{4}{*}{$800$} & Est & 0.801 & 1.007 & 0.599 & 0.301 & 0.254 & 0.494 & 0.506 & 0.547 & 0.749 & 0.756\\
&& SE & 0.130 & 0.132 & 0.033 & 0.046 & 0.036 & 0.044 & 0.058 & 0.219 & 0.276 & 0.124\\
&& $\wh{\mathrm{SE}}$ & 0.125 & 0.129 & 0.032 & 0.042 & 0.034 & 0.042 & 0.057 & 0.206 & 0.239 & 0.121\\
&& CP & 0.938 & 0.944 & 0.952 & 0.926 & 0.932 & 0.938 & 0.952 & 0.918 & 0.902 & 0.944\\

    \bottomrule
\end{tabular}

\end{table}

The PMMH estimator exhibits minimal empirical bias across all aggregation levels, as expected for an estimator that targets the MLE, and 
the approximate $95\%$ confidence intervals are well calibrated. A minor increase in the standard error of the estimates is observed as $\Delta$ increases, reflecting the loss of information. However, the accuracy of the estimates and the estimated standard errors across aggregation levels indicate that the PMMH estimator is robust to heavy aggregation. In general, the biases of the estimates are negligible in comparison to their respective standard errors. Furthermore, the standard errors change by a factor of approximately $1/\sqrt{2}$ when the censoring time is doubled. These behaviours of the PMMH estimator are consistent with those 
of the MLE.  

\subsubsection{Comparison to alternative methods}
\begin{figure}[ht]
\begin{subfigure}{0.49\textwidth}
\centering
    \includegraphics[width = \textwidth]{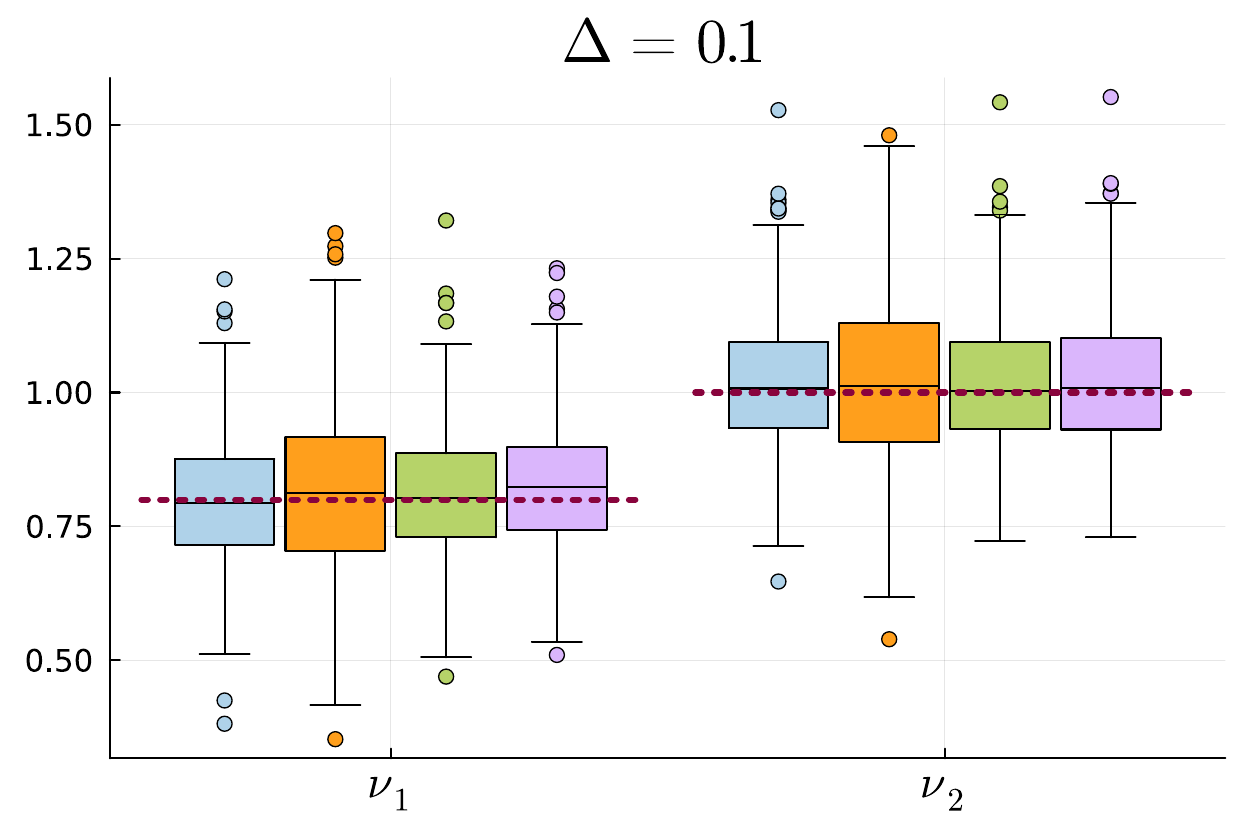}
\end{subfigure}
\hfill
\begin{subfigure}{0.49\textwidth}
\centering
    \includegraphics[width = \textwidth]{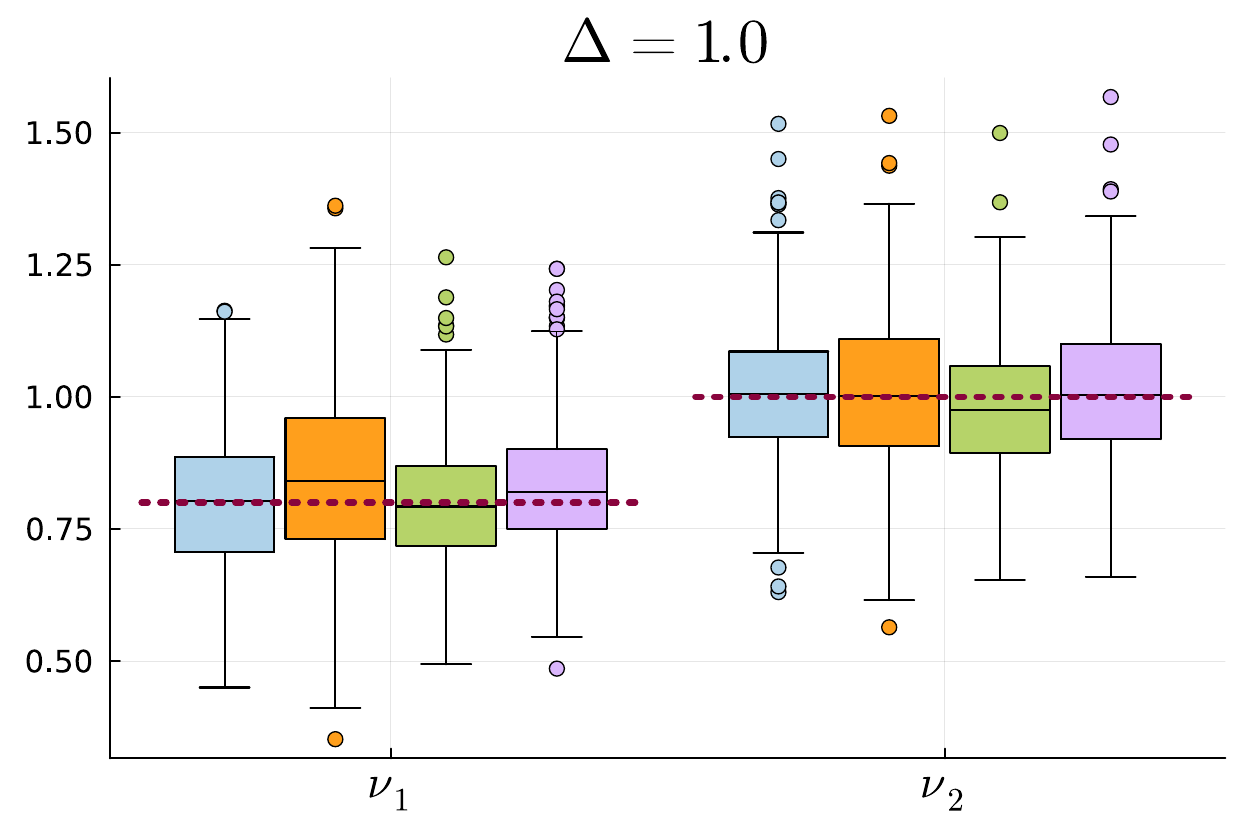}
\end{subfigure}
\\
\begin{subfigure}{0.49\textwidth}
\centering
    \includegraphics[width = \textwidth]{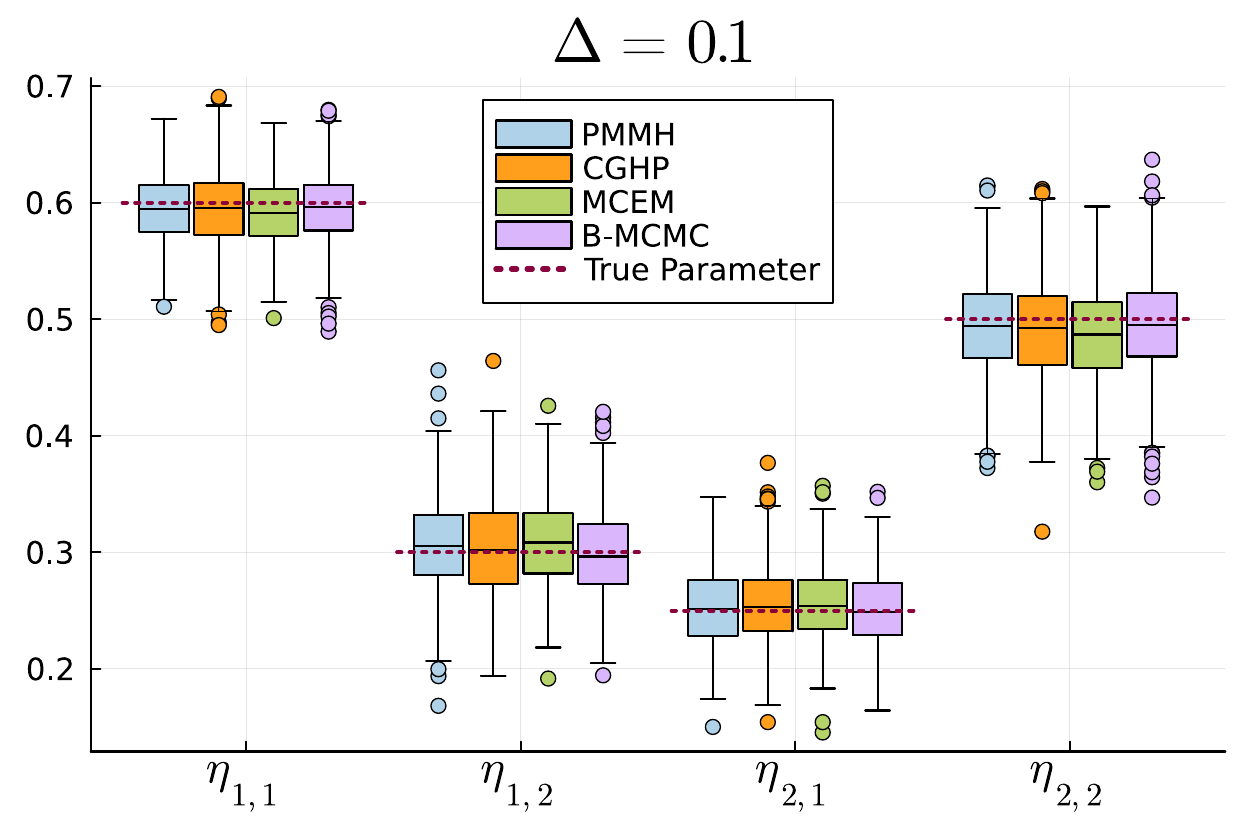}
\end{subfigure}
\hfill
\begin{subfigure}{0.49\textwidth}
\centering
    \includegraphics[width = \textwidth]{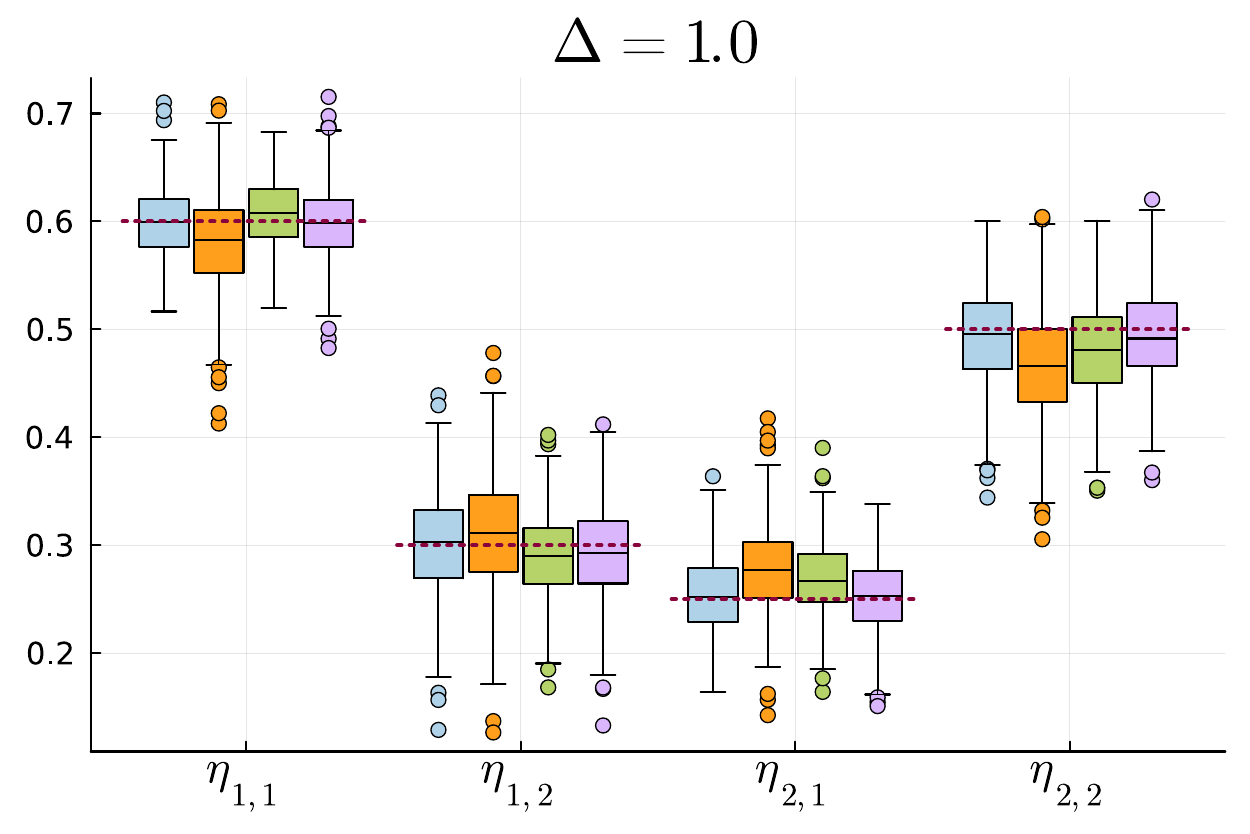}
\end{subfigure}
\\
\begin{subfigure}{0.49\textwidth}
\centering
    \includegraphics[width = \textwidth]{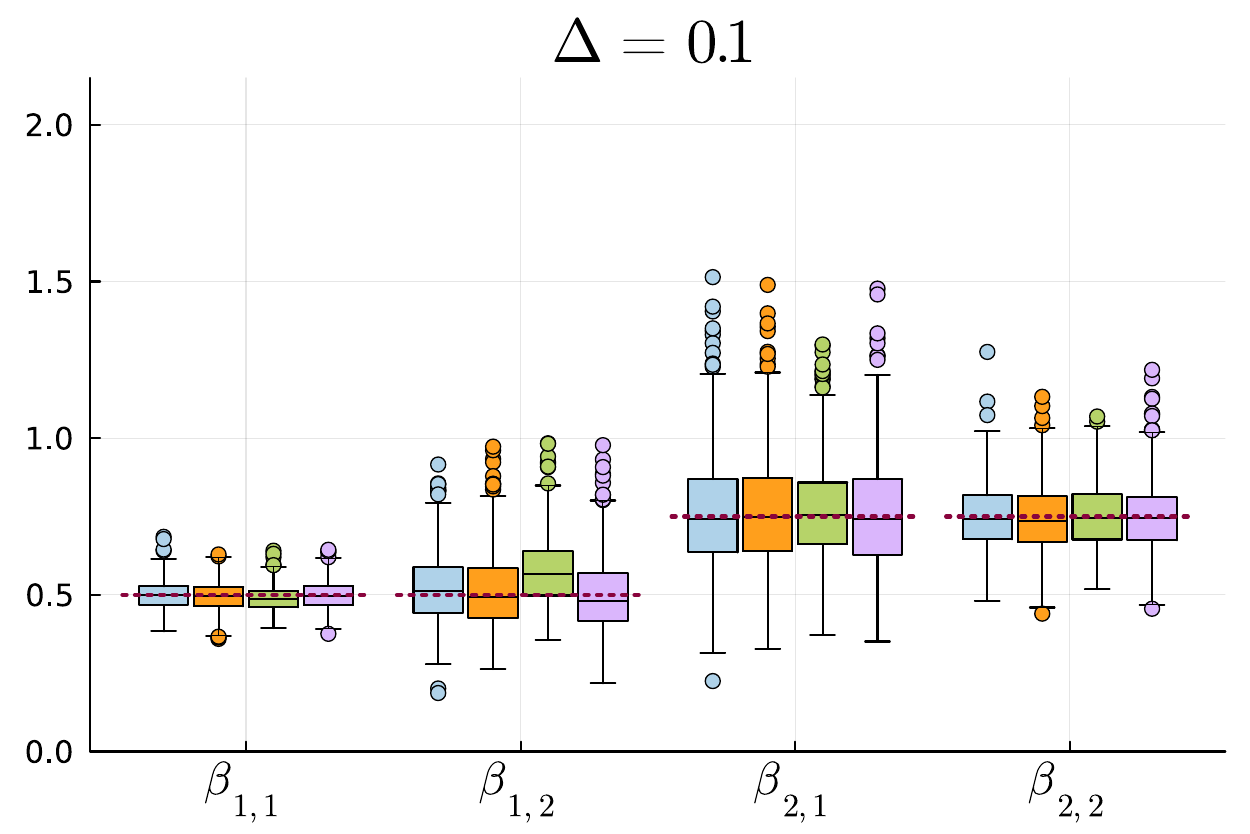}
\end{subfigure}
\hfill
\begin{subfigure}{0.49\textwidth}
\centering
    \includegraphics[width = \textwidth]{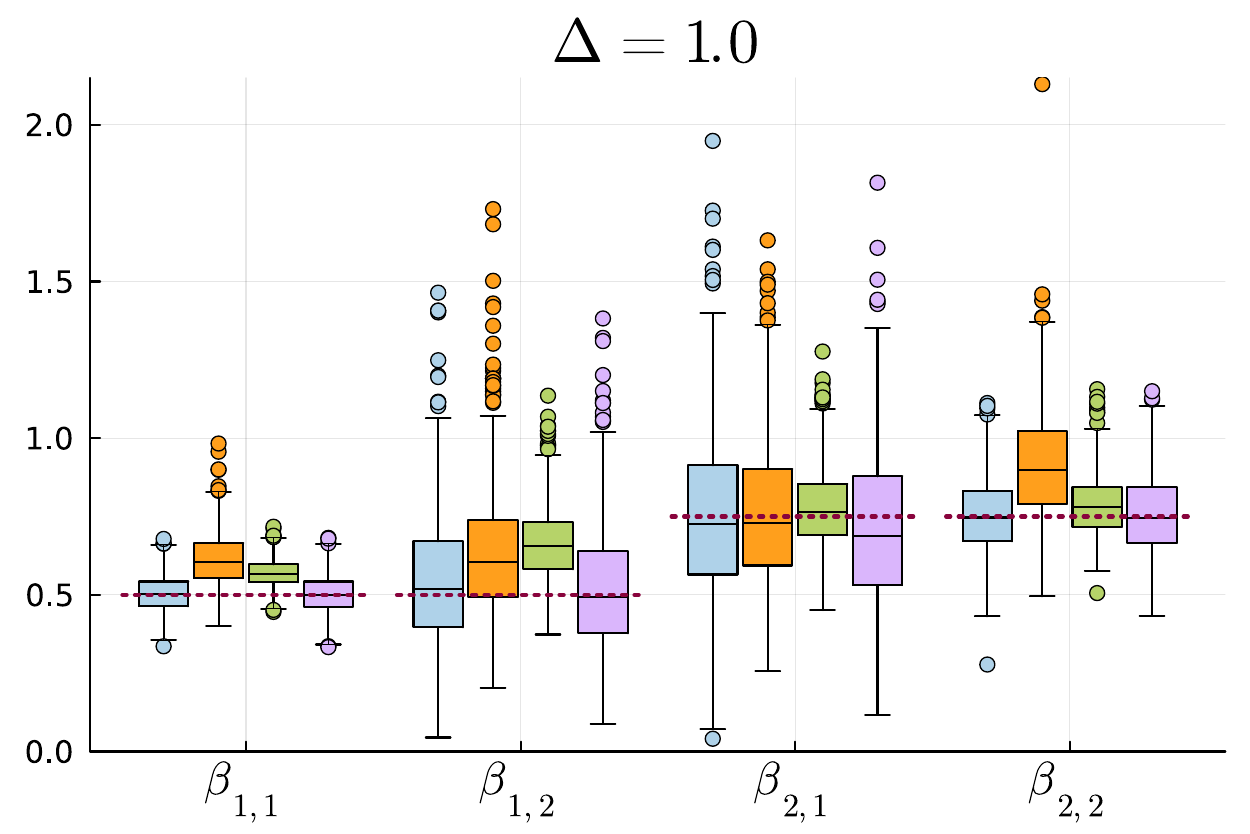}
\end{subfigure}
\caption{Comparison of PMMH, Coarse-Grained Hawkes process (CGHP), MCEM and Bayesian MCMC (B-MCMC) estimators for $T = 800$ and $\Delta\in \{0.1, 1.0\}$.}
\label{fig:method_comp}
\end{figure}

We now present a comparison of the PMMH estimator to some alternative methods in the literature. For the case of $T = 800$ and aggregation widths $\Delta$ presented in Table~\ref{tab:pmmh_sim_exp}, we estimate the parameters using MCEM \citep{shlomovichMultivariate2022}, the coarse-grained Hawkes process (CGHP) estimator \citep{koyamaCoarseGrainedHawkesProcesses2025}, and the Bayesian MCMC estimator \citep{zhouBayesianInferenceAggregated2025}. The MCEM estimates are computed using the \texttt{MATLAB} code released as a supplement to \cite{shlomovichMultivariate2022}\footnote{Available at: 
  \url{https://github.com/lshlomovich/MCEM_Multivariate_Hawkes}
}, using the sequential sampling method, as this is reportedly the least biased of the two proposed techniques. The Bayesian MCMC method is implemented using the supplementary \texttt{R} code to \cite{zhouBayesianInferenceAggregated2025}\footnote{Available at: 
\url{https://github.com/lingxiaozhou/Bayesian-Inference-for-Aggregated-Hawkes-Processes}}. The coarse-grained estimator was implemented in \texttt{Julia} according to the procedure given in \cite{koyamaCoarseGrainedHawkesProcesses2025}. The results are presented in Figure~\ref{fig:method_comp}. 

For the low aggregation case of $\Delta = 0.1$, all methods perform similarly in the estimation of $(\nu_1, \nu_2)$, with somewhat larger variance exhibited by the CGHP estimator due to its use of method of moments estimation for the background parameters. Performance for the estimation of $\eta$ is also similarly good across all three methods, with MCEM exhibiting some significant bias in the estimation of $\beta$. For a larger aggregation level of $\Delta = 1.0$, corresponding to an average of $10.5$ events per censoring interval, significant bias is present in the CGHP and MCEM estimates of $\eta$ and $\beta$. This deterioration in accuracy is expected since for larger $\Delta$, the proposal used in MCEM deviates further from the true distribution, and the CGHP approximation to the true likelihood also becomes less accurate. The Bayesian MCMC estimator exhibits some minor bias in the estimation of $\beta_{2, 1}$ for $\Delta = 1.0$, though performance is generally similar to PMMH in terms of MSE.


For each method, the results were obtained on an Intel Xeon Platinum 8532Y system with 10 CPU cores for PMMH, MCEM, and CGHP. Only one CPU core was used for the Bayesian MCMC estimator as the code provided as a supplement to \cite{zhouBayesianInferenceAggregated2025} does not support parallel computing. For $\Delta = 0.1$, each estimate was produced using $J = 10$ particles and $10,\!000$ iterations of the PMMH chain, requiring an average of $15$ minutes. For MCEM, with $10$ Monte Carlo samples and $10$ iterations of the algorithm (the defaults in the available code), each estimate required an average of $25$ minutes. The CGHP algorithm required approximately, $45$ minutes for estimation since smaller values of $\Delta$ inflate the length of the integer valued time series, making the estimator slower when the aggregation windows are smaller. For PMMH, the total number of events and the number of particles dominate the computational time complexity, so it is not affected by this issue. Each Bayesian MCMC method was run for $3,\!000$ iterations of the chain, averaging $4.5$ hours each. The efficiency of the proposal used for our PMMH algorithm allows for a small number of particles, $J$, making the method faster than the competitors for $\Delta = 0.1$. The PMMH method also scales well with computational resources, as each particle can be updated in parallel, an advantage over Bayesian MCMC.


\section{Analysis of Terror Attack Data}\label{sec:appld_stdy}
The PMMH estimation methodology is used to analyse terror attack data from 2018 to 2021 across Afghanistan and Pakistan. Understanding the patterns of terrorist activity aids the evaluation of terrorism counter-measures; a topic that has long been of interest to criminology researchers \citep{midlarskyWhyViolenceSpreads1980,behlendorfMicrocyclesViolenceEvidence2012,rieber-mohnInvestigationMicrocyclesViolence2021} and policy makers alike \citep{perlCombatingTerrorismChallenge2007,whiteModellingEffectivenessCounterterrorism2014}. Terror attacks are known to exhibit spatiotemporal clustering. For instance, \cite{midlarskyWhyViolenceSpreads1980} found that the spread of international terrorism in Latin America and Western Europe during 1968--1974 exhibited \textit{contagion} effects. More recently, \cite{behlendorfMicrocyclesViolenceEvidence2012} found evidence that terror attacks carried out by the organisations FMLN in El Salvador and ETA in Spain exhibited similar spatiotemporal clustering behaviour, despite differences in history and motive of the two groups. Localised bursts of terrorist activity were also identified in relation to the Taliban insurgency in Afghanistan in 2016 \citep{rieber-mohnInvestigationMicrocyclesViolence2021}. Due to the observed clustering of events, the MHP and other closely related point process models have been applied in the study of terror attack data \citep{porterSelfexcitingHurdleModels2012, tenchSpatiotemporalPatternsIED2016, junFlexibleMultivariateSpatiotemporal2024}. Since terror attack data for a given region are typically reported as daily counts, these works conduct inference by using discretised versions of the Hawkes process or by arbitrarily placing event times within censoring intervals and applying traditional inference methods. The PMMH technique is a sound approach to parameter estimation that does not require ad hoc data or model manipulations. The data for this analysis is obtained from the Global Terrorism Database \citep{gtdGlobalTerrorismDatabase2022}\footnote{Available at the following link: 
  \url{https://www.start.umd.edu/gtd/}
}, a highly comprehensive dataset on worldwide terrorism events.

\subsection{Terrorism in Afghanistan and Pakistan}
The United States of America (US) conducted a complete military withdrawal from Afghanistan over the period of 09 Mar 2020 to 30 Aug 2021 \citep{baldorUSBeginsTroop2020, zeidanWithdrawalUnitedStates2024}. Since this time, Pakistan has reportedly seen a significant increase in terrorist activity, primarily attributed to the \textit{Tehreek-e-Taliban Pakistan} \citep[TTP;][]{akhtarUnderstandingResurgenceTehrikeTaliban2023} and \textit{ISIS Khorosan Province} \citep[IS-KP;][]{AustralianNationalSecurity}. The TTP is an organisation that operates along Pakistan's northwestern border with Afghanistan in the province of Khyber Pakhtunkhwa, which now includes the region formerly called the Federally Administered Tribal Areas \citep{abbasProfileTehrikiTalibanPakistan2008}. The TTP originated as sympathisers to the Afghan Taliban, but since the mid 2000s, they have held their own command structure, independent of the Afghan Taliban \citep{abbasProfileTehrikiTalibanPakistan2008}. Their stated objective is to overthrow the elected government of Pakistan, which they pursue through terror attacks primarily on military and government targets \citep{tehrikUN}. Since the Afghan Taliban assumed control of the Nangarhar province in Afghanistan, the TTP has been afforded safe haven in Afghanistan, resulting in greater capacity for committing acts of terror \citep{ttpUNReport}. IS-KP originated as a splinter group of the TTP in 2014 as sympathisers to the IS leader Abu Bakr al-Baghdadi \citep{AustralianNationalSecurity}. They are often in opposition to the TTP and the Afghan Taliban, rejecting the Afghan Taliban's negotiation with the US, instead aiming for the destruction of all nation states and the establishment of a global caliphate \citep{AustralianNationalSecurity}. IS-KP operates in the provinces of Kabul and Nangarhar in Afghanistan and Khyber Pakhtunkhwa in Pakistan, with the number of attacks committed in these regions by IS-KP growing since the US exit from Afghanistan due to the reduction in US and Afghan government counter-terrorism efforts \citep{AustralianNationalSecurity}. The rise of terrorist activity in Pakistan has garnered significant media attention in recent years, with Pakistan-Afghanistan tensions consistently identified as causal factors in Pakistan's terrorism spike \citep{hussainWhatExplainsDramatic2023, azizTerrorismTensionsPakistan2023, hussainCousinsWarPakistanAfghan2024}.

\begin{figure}
    \centering
    \includegraphics[width=0.75\linewidth]{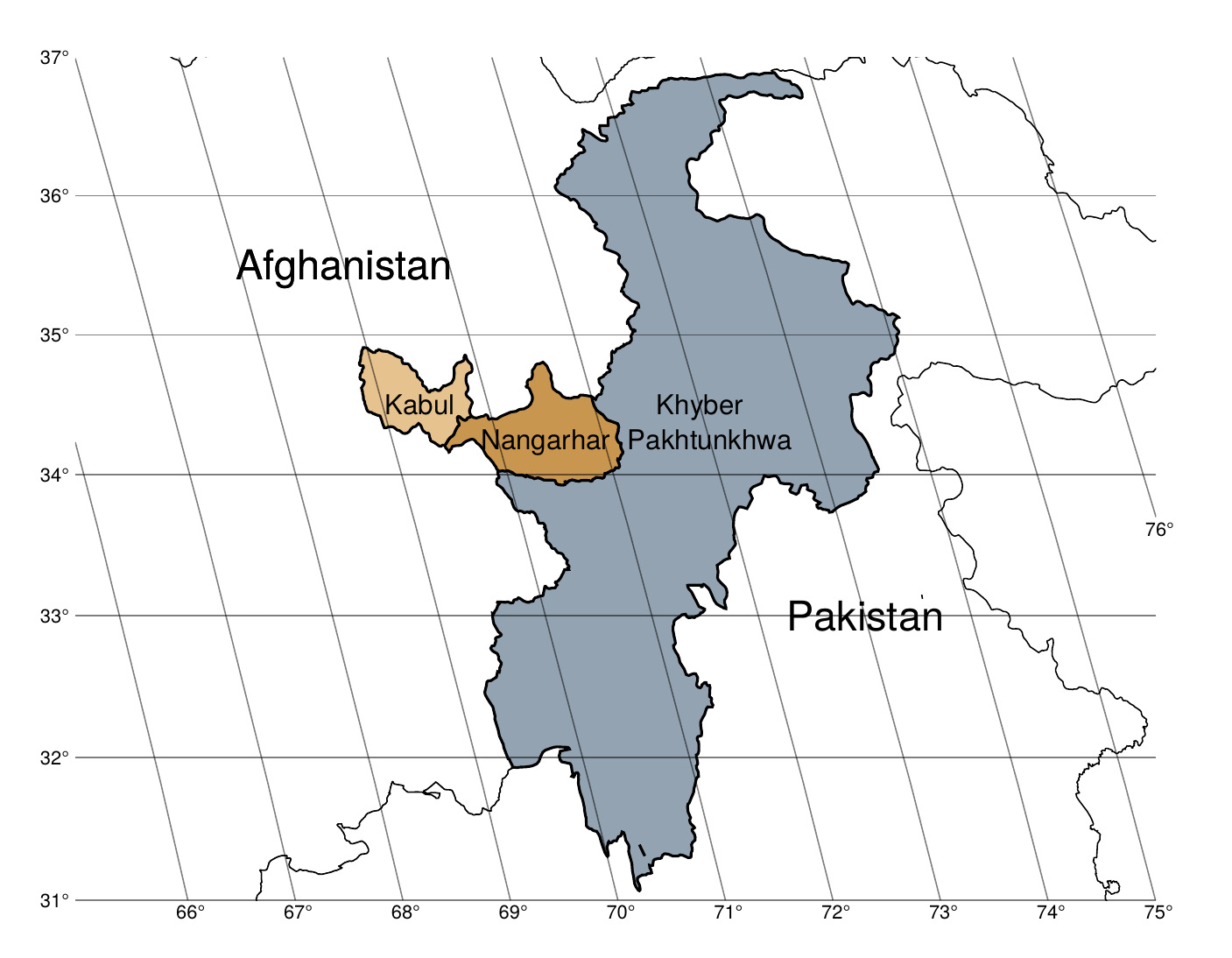}
    \caption{Kabul and Nangarhar provinces of Afghanistan bordering the province of Khyber Pakhtunkhwa in Pakistan.}
    \label{fig:map}
\end{figure}

Given the growth in terrorist activity along the Pakistan-Afghanistan border, we use an MHP to model terror attacks over the period of 01 Jan 2018 to 30 June 2021, which captures a substantial period prior to the US exit from Afghanistan, and the extent of the available data during the exit \citep{gtdGlobalTerrorismDatabase2022}. Specifically, we consider events occurring in the Kabul and Nangarhar provinces of Afghanistan as type~1 events, with events in Khyber Pakhtunkhwa in Pakistan as type~2 events. A map of the regions of interest is displayed in Figure~\ref{fig:map}. Estimates are obtained via our proposed PMMH procedure. The temporal clustering of terrorism in Afghanistan has been studied extensively \citep{zammit-mangionPointProcessModelling2012, rieber-mohnInvestigationMicrocyclesViolence2021, junFlexibleMultivariateSpatiotemporal2024}, suggesting that the nation is an appropriate candidate for a Hawkes process model. To the best of our knowledge, Pakistan has not been the subject of such statistical analysis. Insights into the cross-border dynamics of the two regions during this time period may be drawn from our analysis. For the remainder of this section, we use \enquote{Kabul–Nangarhar} to refer to the region of interest in Afghanistan, emphasising that events in the two provinces are pooled. 

\subsection{Preliminary Analysis}
\begin{figure}[ht]
\centering
\begin{subfigure}{0.49\textwidth}
    \includegraphics[width=\textwidth]{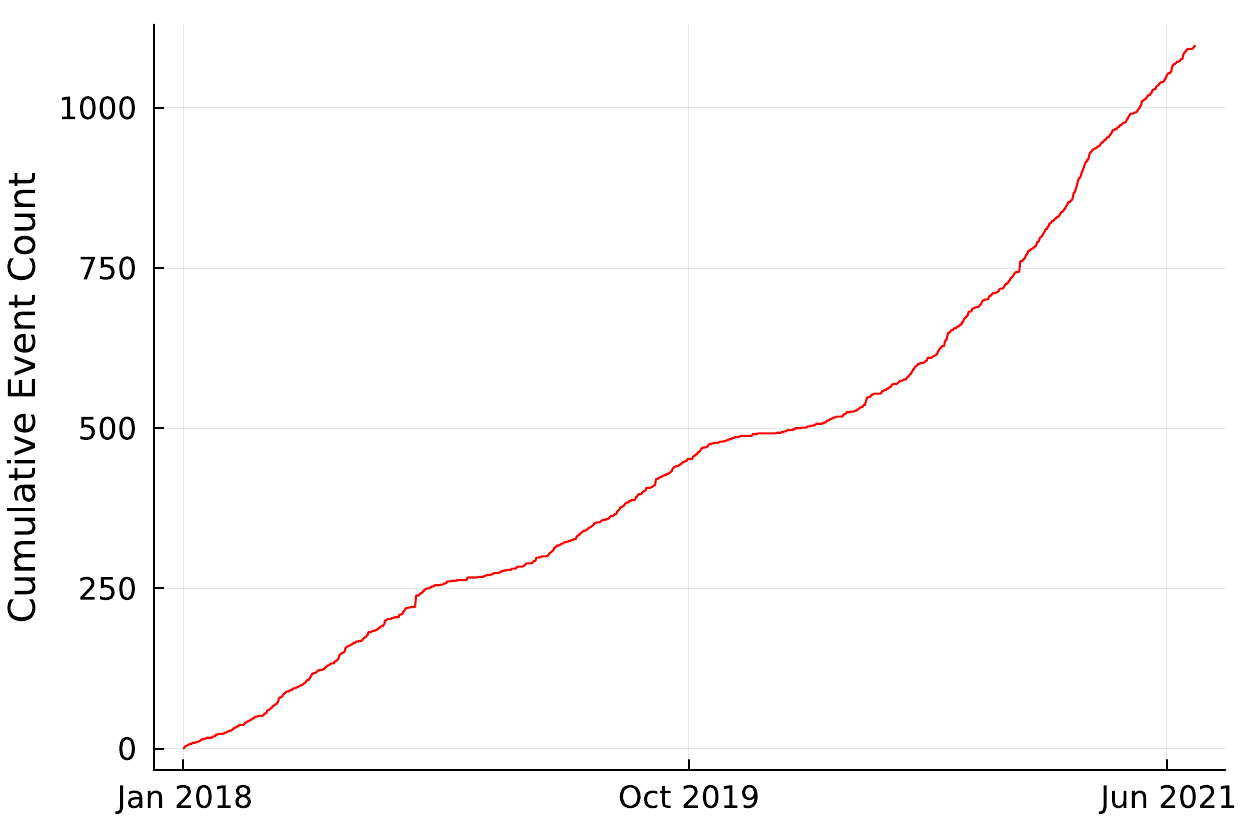}
\end{subfigure}
\hfill
\begin{subfigure}{0.49\textwidth}
    \includegraphics[width=\textwidth]{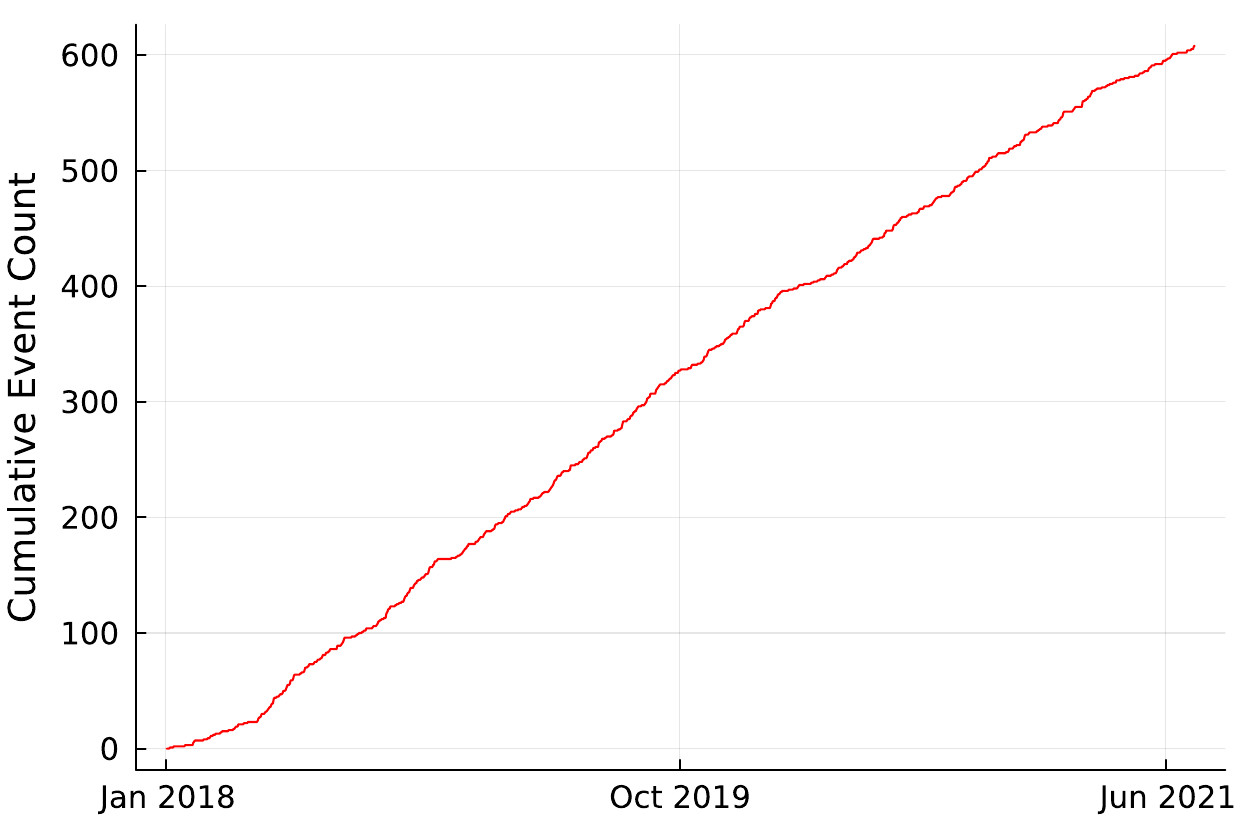}
\end{subfigure}
\\
\begin{subfigure}{0.49\textwidth}
    \includegraphics[width=\textwidth]{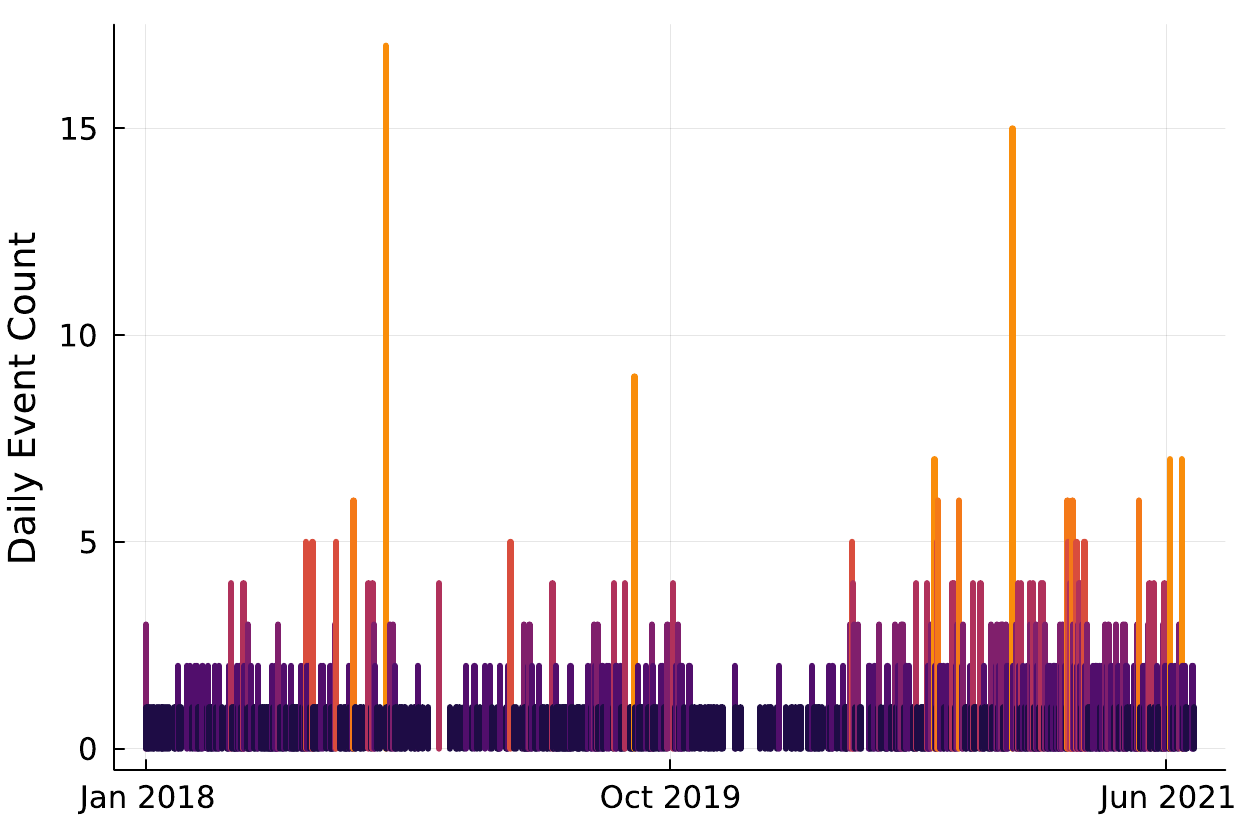}
    \caption{Kabul--Nangarhar}
\end{subfigure}
\hfill
\begin{subfigure}{0.49\textwidth}
    \includegraphics[width=\textwidth]{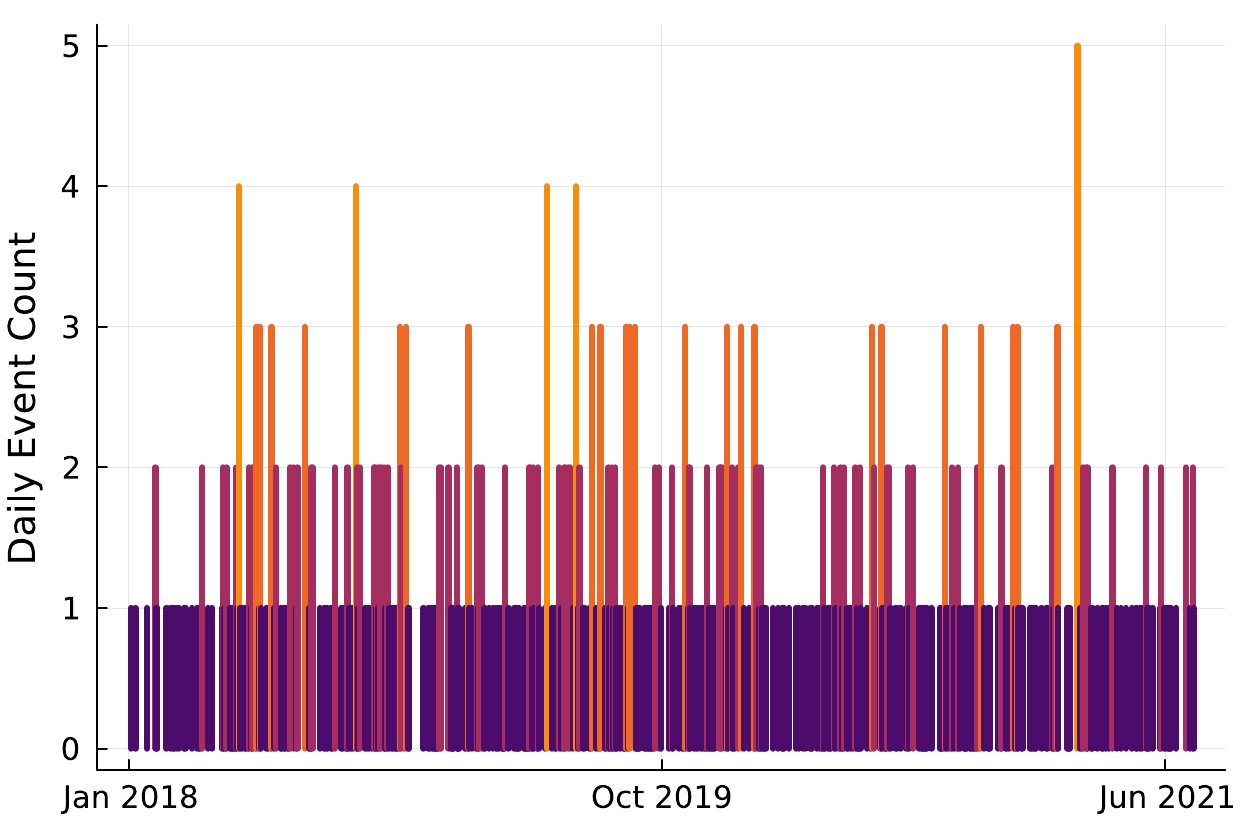}
    \caption{Khyber Pakhtunkhwa}
\end{subfigure}
        
\caption{Top row: Cumulative number of terror attacks in Khyber Pakhtunkhwa and Kabul--Nangarhar, 01 Jan 2018 -- 30 Jun 2021. Bottom row: Daily terror attack counts.}
\label{fig:obs_dat}
\end{figure}

There has been a marked increase in the number of terror attacks in Pakistan and Afghanistan over the period 1970--2021 \citep{gtdGlobalTerrorismDatabase2022}. Both nations experienced few events until the late 1990s and early 2000s, after which an exponential increase in terrorism was observed, attributable either to an actual rise in terrorist activity or a significant increase in reporting of terrorism. Figure~\ref{fig:obs_dat} displays the cumulative event counts for each region of interest over the period 2018--2021. Events in Khyber Pakhtunkhwa appear to occur at a stable rate with only a slight visual tapering, whereas events in Kabul--Nangarhar exhibit more pronounced non-linearity. Figure~\ref{fig:obs_dat} also shows the daily event counts for each region. Kabul--Nangarhar experienced an average of 0.86 events per day, with Khyber Pakhtunkhwa averaging 0.48 daily events. 
Kabul--Nangarhar appears to exhibit more significant temporal spiking than Khyber Pakhtunkhwa, with a maximum of 17 events observed in a single day, while a maximum of only 5 occurred in Khyber Pakhtunkhwa. 

\subsection{Model and Estimation}\label{sec:terr_mod_est}
For a bivariate Hawkes process, $\bs N(t)$, let an event in Kabul--Nangarhar be type 1 and an event in Khyber Pakhtunkhwa be type 2. Setting the initial time $t_0 = 0$ corresponding to 01 Jan 2018, we have a censoring time of $T = 1,\!276$ days, with the $i$\thh observation period defined by $(t_{i-1}, t_i] = (i - 1, i]$. We use a piecewise linear background intensity for events in each region to account for long-term trends in terror attacks. Specifically, $\nu_1(t)$ and $\nu_2(t)$ are order $2$ B-splines with two knots each, located at $\{210, 650\}$ and $\{340, 650\}$ respectively, determined from a visual assessment of the data. The knots correspond to dates 30 July 2018 ($t=210$), 7 Dec 2018 ($t=340$) and 13 Oct 2019 ($t=650$). Following existing works on Hawkes process models of terrorist activity \citep{tenchSpatiotemporalPatternsIED2016, zhouBayesianInferenceAggregated2025}, we adopt an exponential offspring kernel. The parameter $\tht$ is therefore $16$-dimensional.

\begin{figure}[ht]
\centering
\begin{subfigure}{0.49\textwidth}
\centering
    \includegraphics[width = \textwidth]{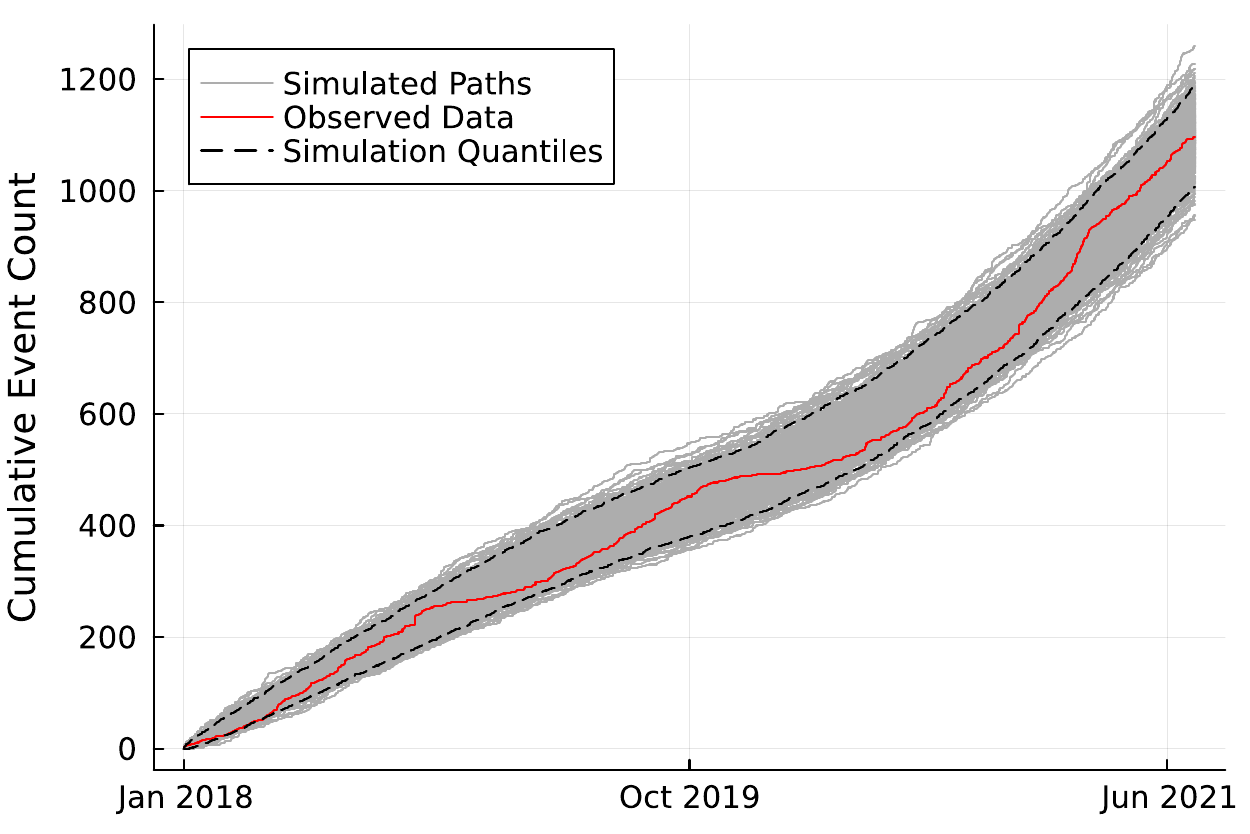}
    \caption{Kabul--Nangarhar}
\end{subfigure}
\hfill
\begin{subfigure}{0.49\textwidth}
\centering
    \includegraphics[width = \textwidth]{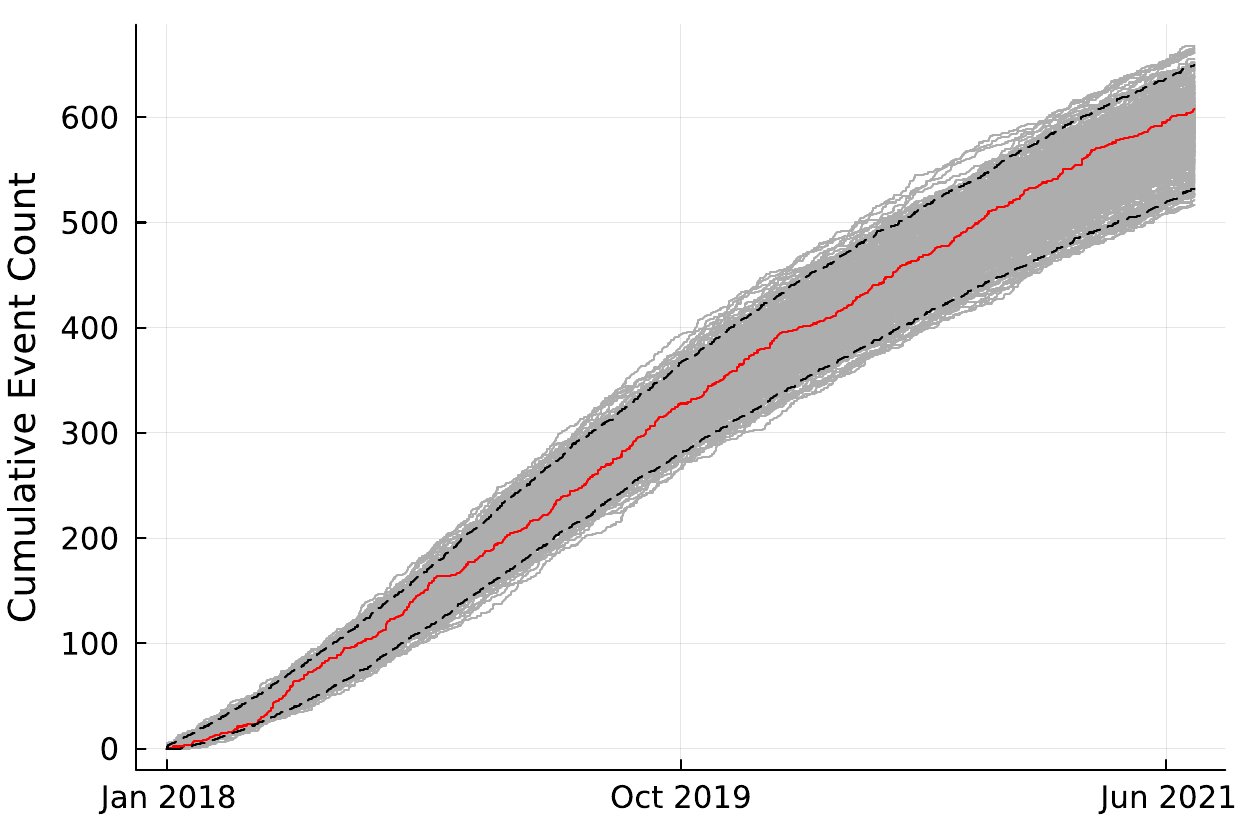}
    \caption{Khyber Pakhtunkhwa}
\end{subfigure}
\caption{Observed cumulative number of terrorist events in Kabul--Nangarhar and Khyber Pakhtunkhwa compared to $500$ paths simulated from $\hth$.}
\label{fig:2018-2021_Sims}
\end{figure}

The PMMH algorithm was run for $40,\!000$ iterations, with a burn-in of $10,\!000$ iterations discarded. Inspection of the trace plots of the MCMC chain shows generally good mixing (Figure~\ref{fig:mcmc}). Histograms of the PMMH chains are also provided (Figure~\ref{fig:terr_hist}). The PMMH estimates and respective standard error estimates are displayed in Table~\ref{tab:2018-2021_summary}. Multiple runs of the PMMH algorithm were performed with different initial parameters, all of which converged to approximately the same location. Figure~\ref{fig:2018-2021_Sims} compares the observed terror attack data across both regions to $500$ sample paths simulated from $\hth$. The observed data falls within the pointwise upper and lower $2.5\%$ quantiles of the simulated paths, suggesting that our estimated model adequately captures the dynamics of terrorism across the two nations over the period of interest.


\begin{table}[ht]
    \centering
    \caption{PMMH estimates and standard error estimates, Kabul--Nangarhar and Khyber Pakhtunkhwa, 2018 -- 2021.}
\label{tab:2018-2021_summary}
    \begin{tabular}{lcccccccc}\toprule
    &$\nu_{1,1}$ & $\nu_{1,2}$ &$\nu_{1,3}$ &$\nu_{1,4}$ & $\nu_{2,1}$ &$\nu_{2,2}$ &$\nu_{2,3}$ & $\nu_{2, 4}$ \\ 
    \midrule
    Est & 0.508 & 0.381 & 0.128 & 1.077 & 0.172 & 0.451 & 0.400 & 0.089\\
$\wh{\mathrm{SE}}$ & 0.096 & 0.122 & 0.092 & 0.114 & 0.059 & 0.065 & 0.049 & 0.077\\
   \toprule
    & $\eta_{1,1}$  & $\eta_{1,2}$ & $\eta_{2,1}$ & $\eta_{2,2}$ & $\beta_{1,1}$ & $\beta_{1,2}$ & $\beta_{2,1}$ & $\beta_{2,2}$\\
   \midrule
   Est & 0.238 & 0.438 & 0.147 & 0.080 & 0.137 & 37.12 & 33.19 & 0.260\\
$\wh{\mathrm{SE}}$ & 0.027 & 0.172 & 0.049 & 0.028 & 0.067 & 15.48 & 23.47 & 0.156\\
    \bottomrule
\end{tabular}

\end{table}

Over the period of 2018--2021, Kabul--Nangarhar saw statistically significant self-excitation effects in terrorist activity, with $\hat\eta_{1, 1}=0.238$ ($\wh{\mathrm{SE}} = 0.027$), which agrees with previous work indicating that terrorism in the Afghanistan occurs in clusters \citep{zammit-mangionPointProcessModelling2012, rieber-mohnInvestigationMicrocyclesViolence2021, junFlexibleMultivariateSpatiotemporal2024}. Self-excitation is also detected in Khyber Pakhtunkhwa ($\hat\eta_{2, 2}=0.08$, $\wh{\mathrm{SE}}=0.028$), though to a lower degree than the border provinces of Afghanistan. Importantly, the estimate $\hat\eta_{1, 2} = 0.438$ ($\wh{\mathrm{SE}} = 0.172$) indicates a great degree of cross-excitation from Khyber Pakhtunkhwa into Kabul--Nangarhar. Terror attacks in Khyber Pakhtunkhwa appear to have a significant triggering effect on subsequent events and Kabul--Nangarhar. A smaller, though still significant, degree of cross excitation from Kabul--Nangarhar into Khyber Pakhtunkhwa is also detected, with $\hat\eta_{2, 1} = 0.147$ ($\wh{\mathrm{SE}} = 0.049$). It is clear from the estimation that terror attacks in the border region of Afghanistan and Pakistan trigger subsequent events across the border. The estimated matrix $\hat\eta$ has spectral radius $\rho(\hat\eta) = 0.42$, reflecting a moderate degree of overall excitation. The estimated matrix $\hat \beta$ suggests that self-excitation in Khyber Pakhtunkhwa and Kabul--Nangarhar occurs over very short time periods, averaging only a few hours, respectively. On the other hand, cross-excitation effects take approximately one month, on average, to occur. This is perhaps due to pre-determined coordination of attacks within a region, creating very short clustering effects, with attacks across regions taking longer to elicit retaliatory or follow-up attacks. Finally, note that the estimated difference $\hat\nu_{1,4} - \hat\nu_{1,3} = 0.95$ is indicative of a significant acceleration in terror attacks in Kabul--Nangarhar in the latter half of the time period in consideration, commensurate with the reduction in counter-terrorism measures in Afghanistan during the period of the US exit \citep{AustralianNationalSecurity}. The estimated background rate for each dimension is displayed in Figure~\ref{fig:bg}.

The specific ideological and political forces causing long-term trends in terror attacks across Pakistan and Afghanistan cannot be discerned from the model. The analysis serves only to identify the degree of excitation across the two regions and the associated timing of the clustering. It is interesting to note that, whilst the regime change in Afghanistan is commonly identified as creating the space for terrorist organisations to operate in Khyber Pakhtunkhwa, the short-term triggering is skewed to flow from Pakistan into Afghanistan. Analysis of an up-to-date dataset will be required to gain insight into whether the dynamics of terrorism across Afghanistan and Pakistan have shifted in the years following the US exit from Afghanistan.

\section{Discussion}\label{sec:conc}
This work contributes a procedure for estimating the parameters of the multivariate Hawkes process from discretely observed data. An unbiased estimate of the intractable likelihood is obtained using SMC with adaptive resampling, with parameter estimates obtained from a PMMH algorithm. Numerical evidence shows that our SMC algorithm, using the ordered uniform proposal distribution, outperforms a multivariate extension of the Poisson proposal in \cite{chenEstimatingHawkesProcess2025}, which enables quality estimates to be obtained with few particles. In simulation experiments, our methodology produces statistically efficient point estimates of the parameters of the MHP across varying levels of aggregation, outperforming the competing MCEM \citep{shlomovichMultivariate2022} and coarse-grained Hawkes process \citep{koyamaCoarseGrainedHawkesProcesses2025} methods in terms of mean-squared error. Performance is comparable to the Bayesian MCMC method of \cite{zhouBayesianInferenceAggregated2025}, although in the examples provided, the PMMH method obtains estimates significantly faster. The PMMH estimator can be used to estimate data from unequally spaced observations and/or with a time-varying background rate, and it provides immediate uncertainty quantification.

In application to terror attack data across Khyber Pakhtunkhwa in Pakistan and Kabul--Nangarhar in Afghanistan, the PMMH estimator produced point estimates that agreed with the observed data. Our method avoids any loss of information from data augmentation used in other Hawkes process models of terror attack data \citep{tenchSpatiotemporalPatternsIED2016, junFlexibleMultivariateSpatiotemporal2024}, and does not require the model to be discretised, as in \cite{porterSelfexcitingHurdleModels2012}.

The estimators drawn from the PMMH algorithm are taken to be the centre of the invariant distribution of the MCMC sample. Although they have demonstrated good performance, we rely fundamentally on the assumption that the MLE is consistent and asymptotically normal to justify this choice. A proof detailing the conditions under which such properties hold will be pursued in other works. Theoretical treatments of SMC and PMMH methods in general contexts are also available \citep{delmoralMeanFieldSimulation2013, andrieuParticleMarkovChain2010}. Limiting results for these algorithms are also of interest for future exploration.

Avenues for improving the computational efficiency of our method remain open. Fine tuning the choice of input parameters to the PMMH algorithm according to the advice in the literature \citep{doucetEfficientImplementationMarkov2015, sherlockEfficiencyPseudomarginalRandom2015} may yield improvements. Recent works have proposed variants of the PMMH algorithm that improve performance in certain contexts, such as \cite{deligiannidisCorrelatedPseudomarginalMethod2018}, which correlate the SMC estimates in subsequent iterations of the MCMC chain to reduce computational time, or \cite{middletonUnbiasedMarkovChain2020}, which utilises coupled Markov chains run in parallel to eliminate burn-in bias. Whether such techniques are beneficial for estimating the discretely observed Hawkes process remains to be seen.




\bibliography{MHP_Bib}

\appendix

\section{Derivations}\label{app:A}
\subsection{Importance Weights}\label{app:A1_lw}
We now derive an explicit representation of the integral of $\lms(t)$, which is needed when computing the potential functions $G_i$ in \eqref{eq:G}. Let $\Psi_{m, k}(t) = \int_0^t\psi_{m,k}(s)\dd s$ for all $m,k\in \cc M$, and let $V_i = \int^{t_i}_{t_{i-1}}\nu(s)\dd s$. The integral is
\begin{align*}
	\int_{t_{i-1}}^{t_i}\lms(s)ds \ &= \ V_i \ + \ \sum_{m=1}^M \int_{t_{i-1}}^{t_i}\vp_m(s)ds.
\end{align*}
For any given $m\in \cc M$, write
\begin{align}
	 \int_{t_{i-1}}^{t_i}\vp_m(s)ds \ &= \ \int_{t_{i-1}}^{\tau_{\Ns_{i-1}+1}}\vp_m(s)ds \ + \ \sum_{K=\Ns_{i-1}+2}^{\Ns_i}\int_{\tau_{K-1}}^{\tau_K}\vp_m(s)ds\nonumber \\ &\hspace{3cm}+ \ \int^{t_i}_{\tau_{\Ns_i}}\vp_m(s)ds.\label{eq:integ_phim}
\end{align}
The first integral is
\begin{align}
	\int_{t_{i-1}}^{\tau_{\Ns_{i-1}+1}}\vp_m(s)ds \ &= \ \sum_{k=1}^{\Ns_{i-1}}\int_{t_{i-1}}^{\tau_{\Ns_{i-1}+1}}\psi_{m,z_k}(s \, - \, \tau_k)ds\nonumber\\ 
	&= \ \sum_{k=1}^{\Ns_{i-1}}\big\{\Psi_{m,z_k}(\tau_{\Ns_{i-1}+1} \, - \, \tau_k) \ - \ \Psi_{m,z_k}(t_{i-1} \, - \, \tau_k)\big\},\label{eq:G_sum1}
\end{align}
with the final integral similarly being
\begin{align}
    \int^{t_i}_{\tau_{\Ns_i}}\vp_m(s)ds \ &= \ \sum_{k=1}^{\Ns_i}\big\{\Psi_{m,z_k}(t_i \, - \, \tau_k) \ - \ \Psi_{m,z_k}(\tau_{\Ns_i} \, - \, \tau_k)\big\}\nonumber \\
    &= \ \sum_{k=1}^{\Ns_i}\Psi_{m,z_k}(t_i \, - \, \tau_k) \ - \ \sum_{k=1}^{\Ns_{i-1}}\Psi_{m,z_k}(\tau_{\Ns_i} \, - \, \tau_k). \label{eq:G_sum2}
\end{align}
The integrals between event times in \eqref{eq:integ_phim} are given by
\begin{align*}
	 &\sum_{K=\Ns_{i-1}+2}^{\Ns_i}\int_{\tau_{K-1}}^{\tau_K}\vp_m(s)ds\ =  \sum_{K=\Ns_{i-1}+2}^{\Ns_i}\sum_{k=1}^{K-1}\int_{\tau_{K-1}}^{\tau_K}\psi_{m, z_k}(s\, -\, \tau_k)ds\\
	 &\qquad =  \sum_{K=\Ns_{i-1}+2}^{\Ns_i}\sum_{k=1}^{K-1}\Psi_{m, z_k}(\tau_K\, -\, \tau_k)\ -\ \sum_{K=\Ns_{i-1}+2}^{\Ns_i}\sum_{k=1}^{K-1}\Psi_{m, z_k}(\tau_{K-1}\, -\, \tau_k)\\
	 &\qquad=  \sum_{K=\Ns_{i-1}+2}^{\Ns_i}\sum_{k=1}^{K-1}\Psi_{m, z_k}(\tau_K\, -\, \tau_k)\ - \ \sum_{K=\Ns_{i-1}+1}^{\Ns_i-1}\sum_{k=1}^{K}\Psi_{m, z_k}(\tau_{K}\, -\, \tau_k),
\end{align*}
where for the final equality we have simply relabelled the index $K$ on the second summation term. To continue, note that since the offspring density $h_{m,k}(\cdot)$ is for a continuous random variable with non-negative support, $\Psi_{m,k}(0) = 0$. Therefore,
\begin{align}
	&\sum_{K=\Ns_{i-1}+2}^{\Ns_i}\sum_{k=1}^{K-1}\Psi_{m, z_k}(\tau_K\, -\, \tau_k)\ - \ \sum_{K=\Ns_{i-1}+1}^{\Ns_i-1}\sum_{k=1}^{K}\Psi_{m, z_k}(\tau_{K}\, -\, \tau_k)\nonumber\\
	&\hspace{2cm} = \ \sum_{k=1}^{\Ns_i - 1}\Psi_{m, z_k}(\tau_{\Ns_i}\, -\, \tau_k) \ - \ \sum_{k=1}^{\Ns_{i-1}}\Psi_{m, z_k}(\tau_{\Ns_{i-1}+1}\, -\, \tau_k).\label{eq:G_sum3}
\end{align}
Combining \eqref{eq:G_sum1}, \eqref{eq:G_sum2} and \eqref{eq:G_sum3} gives
\begin{align}
	\int_{t_{i-1}}^{t_i}\vp_m(s)ds \ &= \ \sum_{k=1}^{\Ns_i}\Psi_{m, z_k}(t_i\, -\, \tau_k) \ - \ \sum_{k=1}^{\Ns_{i-1}}\Psi_{m,z_k}(t_{i-1} \, - \, \tau_k).\label{eq:integ_phim_final}
\end{align}
Summing over $m\in \cc M$ gives the desired integral. Exploiting the many cancellations in this derivation yields dramatic computational speed improvements when compared to computing every term.


\subsection{SMC with Exponential Kernel}\label{app:A2_exp}
For an MHP specified with exponential kernels, the intensity of type-$m$ events is given by
\begin{align*}
    \lm_m(t) \ &= \ \nu_m(t) \ + \ \sum_{\tau_k < t}\frac{\eta_{m,z_k}}{\beta_{m,z_k}}\exp\Big(-\frac{t\, -\, \tau_k}{\beta_{m,z_k}}\Big).
\end{align*}
Define the stochastic matrix $\vep(t)\in \bb R_+^{M\times M}$ to have entries
\begin{align*}
    \vep_{m,p}(t) \ &= \ \sum_{\tau_k < t}\frac{\eta_{m,z_k}}{\beta_{m,z_k}}\exp\Big(-\frac{t\, -\, \tau_k}{\beta_{m,z_k}}\Big)\ind_{\{p\}}(z_k) \\ &= \ \sum_{\tau_k < t}\frac{\eta_{m,p}}{\beta_{m,p}}\exp\Big(-\frac{t\, -\, \tau_k}{\beta_{m,p}}\Big)\ind_{\{p\}} (z_k).
\end{align*}
The entry $\vep_{m,p}(t)$ contains the contribution of type-$p$ events to $\lm_m(t)$. Suppose that, on the $i$\thh observation interval, events $t_{i-1}\, <\, \tau_{\Ns_{i-1}+1}\, <\, \dots\, <\, \tau_{\Ns_i}\, \leq\, t_i$ are observed. On $(t_{i-1},\, \tau_{\Ns_{i-1}+1}]$, we have
\begin{align*}
    \vep_{m,p}(t) \ &= \ \sum_{\tau_k < t}\frac{\eta_{m,p}}{\beta_{m,p}}\exp\Big(-\frac{t_{i-1}\, -\, \tau_k}{\beta_{m,p}}\Big) \exp\Big(-\frac{t\, -\, t_{i-1}}{\beta_{m,p}}\Big)\ind_{\{p\}}(z_k) \\
    &= \ \vep_{m,p}(t_{i-1})\exp\Big(-\frac{t\, -\, t_{i-1}}{\beta_{m,p}}\Big).
\end{align*}
For $K\, =\, \Ns_{i-1}+2,\, \dots,\, \Ns_i$, if $t\in (\tau_{K-1},\, \tau_K]$ then
\begin{align*}
    \vep_{m,p}(t) \ &= \ \sum_{\tau_k < t}\frac{\eta_{m,p}}{\beta_{m,p}}\exp\Big(-\frac{\tau_{K-1}\, -\, \tau_k}{\beta_{m,p}}\Big) \exp\Big(-\frac{t\, -\, \tau_{K-1}}{\beta_{m,p}}\Big)\ind_{\{p\}}(z_k) \\
    &= \ \Big(\vep_{m,p}(\tau_{K-1})\ +\ \frac{\eta_{m,p}}{\beta_{m,p}}\ind_{\{p\}}(z_{K-1})\Big)\cdot\exp\Big(-\frac{t\, -\, \tau_{K-1}}{\beta_{m,p}}\Big).
\end{align*}
The preceding expression shows that, for all $m,\, p$, the value $\vep_{m,p}(\tau_{K-1})$ decays on the increment $(\tau_{K-1},\, t]$, but that the additional jump $\eta_{m,p}/\beta_{m,p}$ appears if the event at $\tau_{K-1}$ is of type $z_{K-1} = p$. Finally, on $(\tau_{\Ns_i},\, t_i]$, we similarly have
\begin{align*}
    \vep_{m,p}(t) \ &= \ \Big(\vep_{m,p}(\tau_{\Ns_i})\ +\ \frac{\eta_{m,p}}{\beta_{m,p}}\ind_{\{p\}}(z_{\Ns_i})\Big)\cdot\exp\Big(-\frac{t\, -\, \tau_{\Ns_i}}{\beta_{m,p}}\Big).
\end{align*}
For completion, note that if no event is present on the $(t_{i-1},\, t_i]$, then for $t\in (t_{i-1},\, t_i]$,
\begin{align*}
    \vep_{m,p}(t) \ &= \ \vep_{m,p}(t_{i-1})\exp\Big(-\frac{t\, -\, t_{i-1}}{\beta_{m,p}}\Big).
\end{align*}
With these values in hand, we can compute the integral present in \eqref{eq:G}. Define $\cc E(\cdot;\,\beta)$ to be the CDF of an exponential random variable with parameter $\beta$. Then
\begin{align*}
    \int_{t_{i-1}}^{\tau_{\Ns_{i-1}+1}}\vep_{m,p}(t)dt \ &= \ \vep_{m,p}(t_{i-1})\, \beta_{m,p}\, \cc E(\tau_{\Ns_{i-1}+1}\, -\, t_{i-1};\, \beta_{m,p}).
\end{align*}
For $K\, =\, \Ns_{i-1}+2,\, \dots,\, \Ns_i$,
\begin{align*}
    \int_{\tau_{K-1}}^{\tau_{K}}\vep_{m,p}(t)dt \ &= \ \Big(\vep_{m,p}(\tau_{K-1})\ +\ \frac{\eta_{m,p}}{\beta_{m,p}}\ind_{\{p\}}(z_k)\Big)\, \beta_{m,p}\, \cc E(\tau_{K}\, -\, \tau_{K-1};\, \beta_{m,p}).
\end{align*}
A similar expression is available for the integral of $\varepsilon_{m,p}(s)$ over $(\tau_{\Ns_i}, t_i]$.
Note that
\begin{align}\label{eq:exp_integ}
    \int_{t_{i-1}}^{t_i}\lms(s)\dd s \ &= \ \int_{t_{i-1}}^{\tau_{\Ns_{i-1}+1}}\lm(s)\dd s \ + \ \sum_{K= \Ns_{i-1}+2}^{\Ns_i}\int_{\tau_{K-1}}^{\tau_K}\lm(s)\dd s\nonumber\\
    &\hspace{3cm} + \ \int^{t_i}_{\tau_{\Ns_i}}\lm(s)\dd s.
\end{align}
Writing $\lms(s) = \nu(s) \ + \ \sum_{m,p=1}^M \vep_{m,p}(s)$, the integrals in \eqref{eq:exp_integ} are easily calculated by summing over the integrals previously derived. Finally, consider the sum $\sum_{k=\Ns_{i-1}+1}^{\Ns_i} \log \lm_{z_k}(\tau_k)$,
which is present in the expression for $\log G_i$. We have that
\begin{align*}
    \lm_{z_k}(\tau_k) \ &= \ \nu_{z_k}(\tau_k) \ + \ \sum_{p=1}^M\vep_{z_k,p}(\tau_k),
\end{align*}
hence the sum in question can be computed incrementally using the matrix $\vep(t)$. The advantage of the exponential case is that, through the representations provided, relevant values of the self-excitation effect $\vep(t)$ can be calculated recursively. This is of linear time complexity, and eliminates the need to store the complete history of a particle chain.

\section{Applied Study}\label{app:C}
\begin{figure}[h!]
    \centering
    \includegraphics[width=1\linewidth]{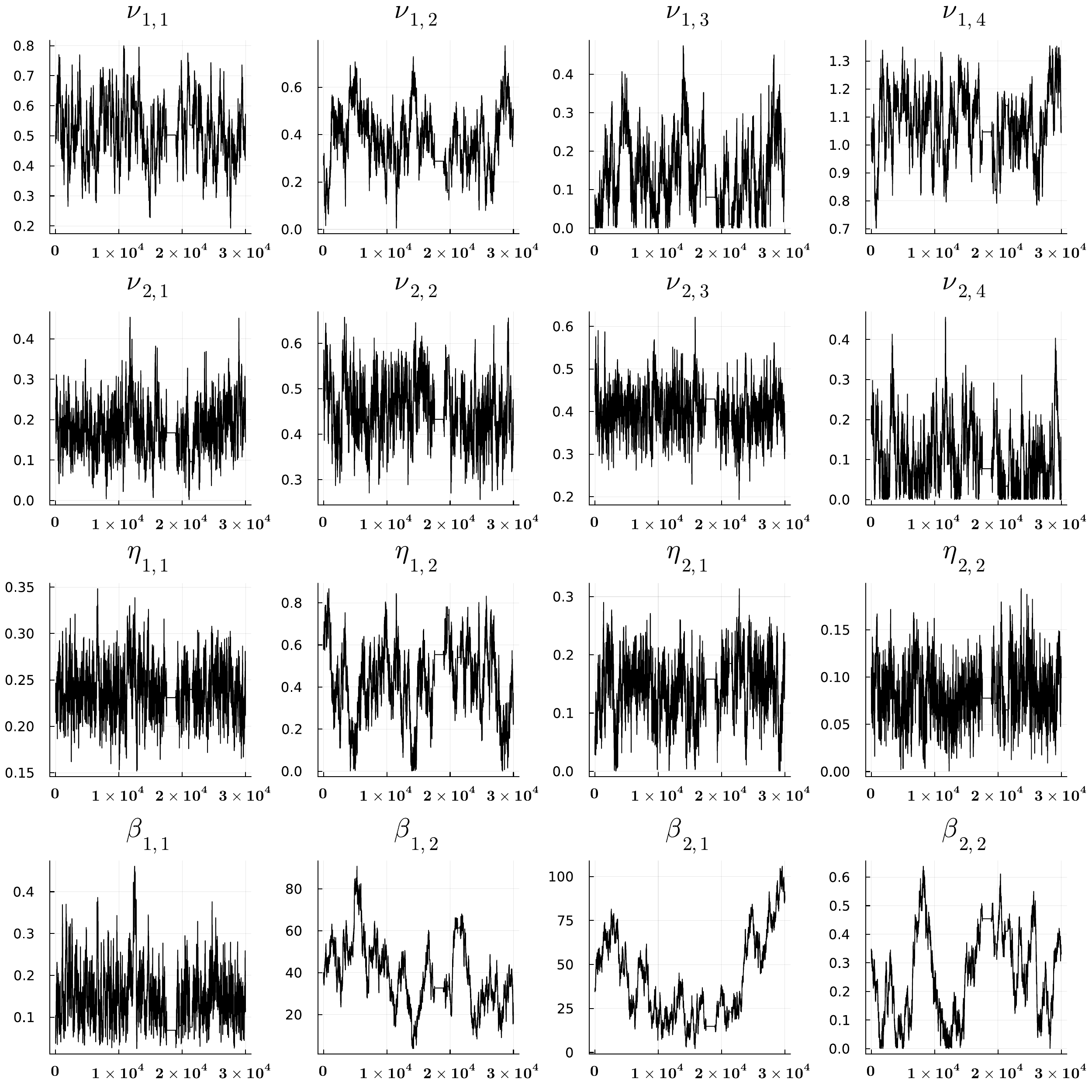}
    \caption{Trace plots for PMMH chains associated with the estimates in Table~\ref{tab:2018-2021_summary}}
    \label{fig:mcmc}
\end{figure}

\begin{figure}[h!]
    \centering
    \includegraphics[width=1\linewidth]{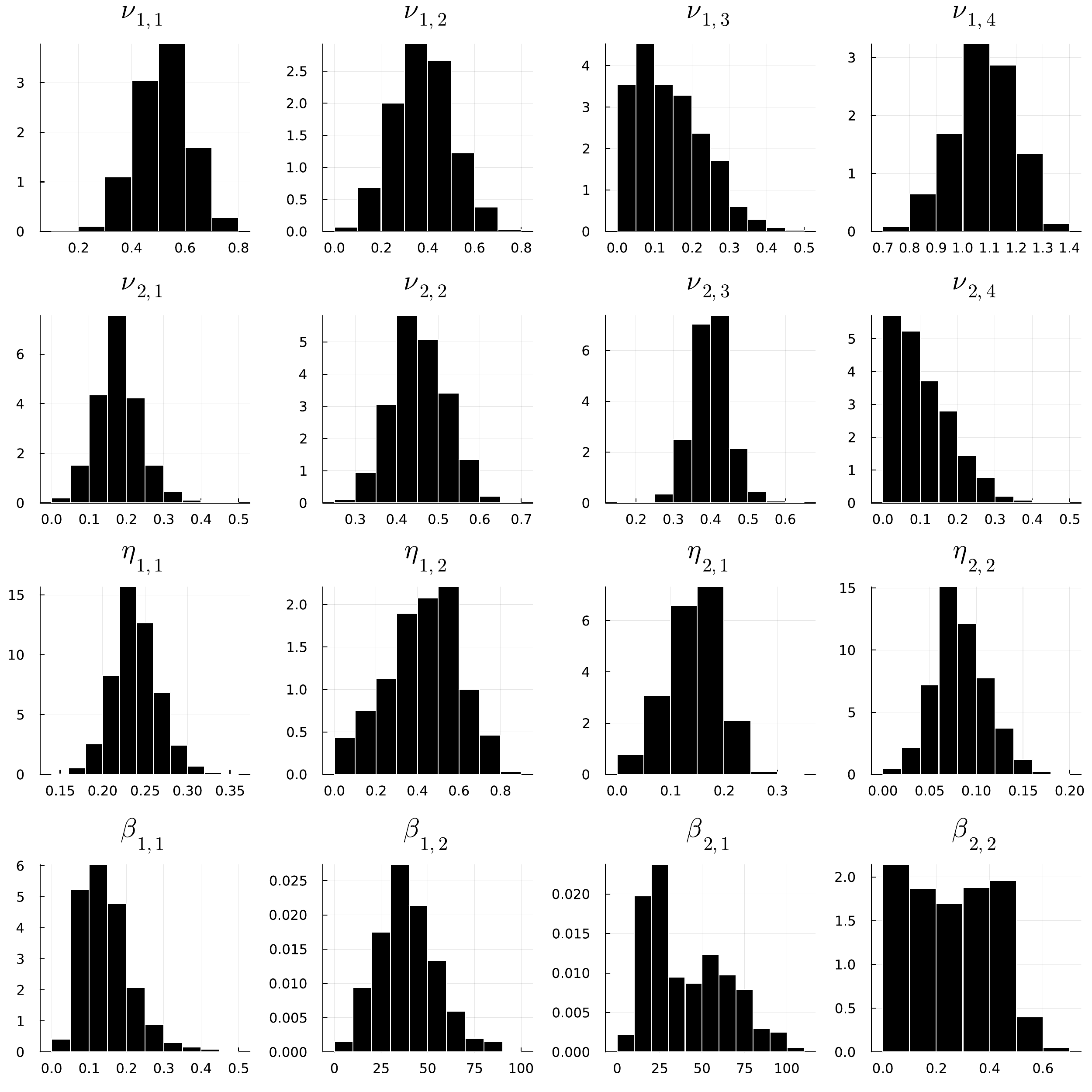}
    \caption{Histograms of PMMH chains associated with the estimates in Table~\ref{tab:2018-2021_summary}}
    \label{fig:terr_hist}
\end{figure}

Figure~\ref{fig:mcmc} shows the PMMH algorithm trace plots for each parameter estimate in Table~\ref{tab:2018-2021_summary}. We observe good mixing of the chain across the background rate and branching ratio parameters. Figure~\ref{fig:terr_hist} displays the same data in the form of histograms, from which we observe approximate convergence with the normal distribution, as expected. Note that certain dimensions of the background parameter (top two rows, Figure~\ref{fig:terr_hist}) are skewed due to the proximity of the parameter to the $0$ lower bound. Additionally, the histograms associated with $\hat\beta_{2, 1}$ and $\hat{\beta}_{2, 2}$ are distorted due to the low values of $\hat\eta_{1, 2}$ and $\hat\eta_{2, 2}$, respectively, making the offspring density challenging to identify along these dimensions.

\begin{figure}[h]
\centering
\begin{subfigure}{0.49\textwidth}
\centering
    \includegraphics[width = \textwidth]{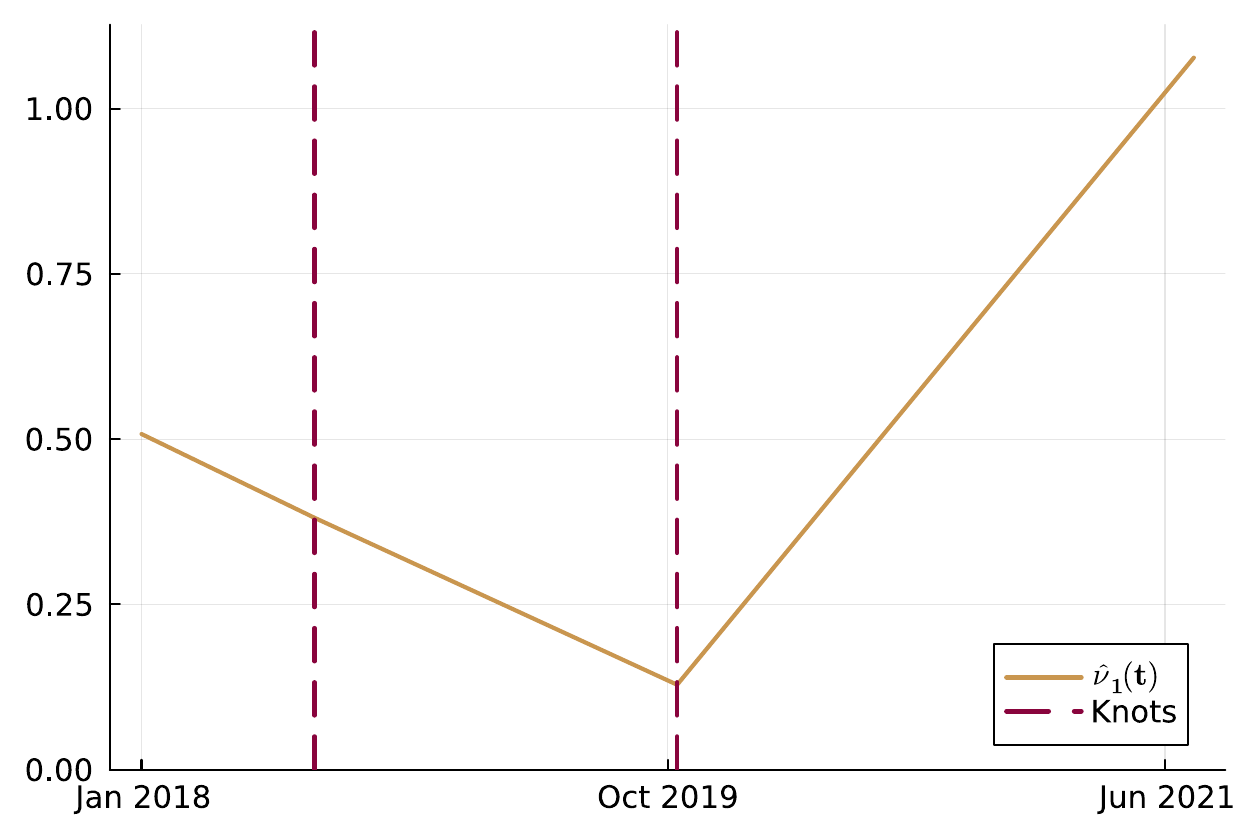}
    \caption{Kabul--Nangarhar}
\end{subfigure}
\hfill
\begin{subfigure}{0.49\textwidth}
\centering
    \includegraphics[width = \textwidth]{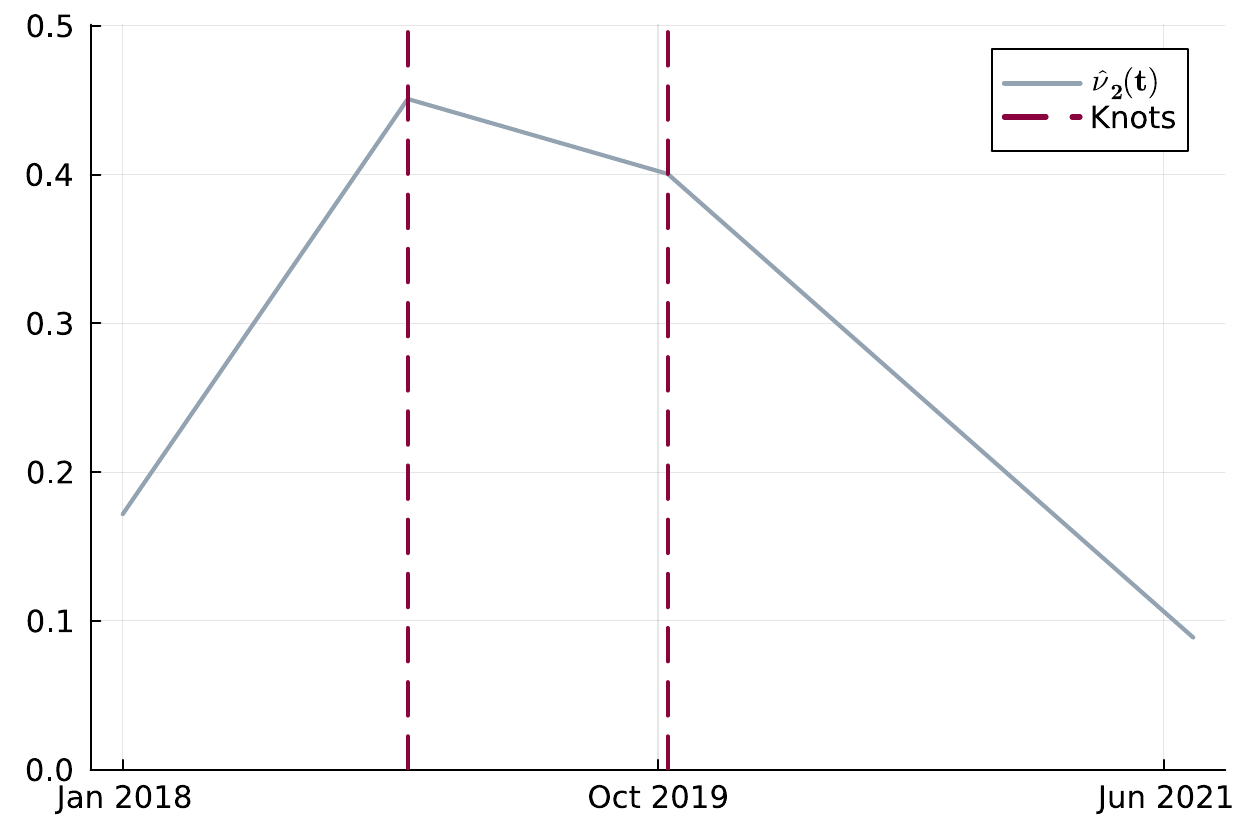}
    \caption{Khyber Pakhtunkhwa}
\end{subfigure}
\caption{Estimated background intensity functions.}
\label{fig:bg}
\end{figure}

Figure~\ref{fig:bg} displays the estimated background rate functions for Kabul--Nangarhar and Khyber Pakhtunkhwa, respectively. For Kabul--Nangarhar we see a slight decrease in background intensity until the second knot, after which there is a large increase in estimated background intensity. The estimated background rate in Khyber-Pakhtunkhwa instead suggests a decrease in background events in the later part of the period of interest.

\end{document}